\documentclass[11pt]{article}
\pdfoutput=1
\usepackage{times}
\usepackage{graphics}
\usepackage{graphicx}
\usepackage{latexsym}
\usepackage{amssymb,amsmath,amsthm}
\usepackage{harvard,ulem}
\usepackage{epstopdf}
\usepackage{enumerate}
\usepackage{bm}
\usepackage{setspace}
\usepackage{fullpage}
\usepackage{sectsty,bbding,lscape}
\usepackage{times}
\usepackage{lscape}
\usepackage{anysize}
\usepackage{fancyhdr,placeins,graphicx,tikz,tikz-layers}
\usetikzlibrary{positioning,calc,arrows,arrows.meta, intersections,shapes.geometric,shadows}
\usepackage{booktabs}       % For professional-quality table lines
\usepackage{amsmath}        % For math symbols like \hat{} and subscripts
\usepackage{threeparttable}

\numberwithin{equation}{section}
\newtheorem{proposition1}{Proposition}[section]

\newtheorem{definition1}{Definition}[section]
\newtheorem{remark1}{Remark}[section]

\newtheorem{lemma1}{Lemma}[section]

\newtheorem{corollary1}{Corollary}[section]
\newtheorem{example1}{Example}[section]
\begin{document}

\begin{titlepage}
\title{{\sc  Random Disclosure Costs: A Unified Theory of Disclosure}}
\author{
Jeremy Bertomeu, \textit{Washington University in St. Louis} \\
Edwige Cheynel, \textit{Washington University in St. Louis} \\
Keri Hu, \textit{The University of Hong Kong}\footnote{
Contact Author: Jeremy Bertomeu, Professor, Olin Business School, Washington University in St. Louis, One Brookings Drive, St. Louis, MO 63130-4899. 
Phone: 314-935-6000. Email: b(firstname)@wustl.edu. 
The authors thank Davide Cianciaruso, Mario Milone, Iv\'{a}n Marinovic, Kevin Smith, and many Ph.D. students from Jeremy Bertomeu's theory foundations course for valuable advice at various stages of this paper.
}
}
\date{}
\maketitle
\begin{abstract}
We propose a unified framework for a class of evidence games in which a sender, seeking to increase posterior expectations (e.g., a market price), chooses whether to reveal or withhold her information. Our model nests several classic models, including type-dependent disclosure costs or benefits, uncertain information endowment, and the market for lemons. Frictions are captured as random disclosure costs that are privately known to the sender and noisily inferred by the receiver. We derive a simple characterization of the equilibrium and show that it features increasing disclosure for lower frictions or greater fundamental uncertainty.
Further, with high uncertainty or low frictions, greater uncertainty \textit{about disclosure costs} decreases the likelihood of disclosure. Positive costs do not necessarily rule out unraveling to full disclosure, especially when outcomes have heavy tails or when disclosure costs are not constant. Finally, we demonstrate that the model can be estimated with a simple probit, suggesting that cost uncertainty is descriptive of real-world behavior.
\vspace{0.3in}
\vspace{0in}

\bigskip
\end{abstract}
\setcounter{page}{0}
\thispagestyle{empty}
\end{titlepage}
%-Title

\begin{singlespace}

\section{Introduction}

\paragraph{}

%\onehalfspacing
%\addtolength{\parskip}{-5 mm}

%\pdfgraphics

Existing literature has given rise to two seemingly antagonistic approaches to modeling the frictions to the disclosure of evidence. The first approach, initiated by \citeasnoun{jov82} and \citeasnoun{ver83}, posits a disclosure cost, such that the informed party incurs a loss from publicly disclosing information. Such cost is typically interpreted as arising from the (implicit) verification needed for the disclosure to be credible, or  
from the potential adverse consequences of disclosure on future interactions with third parties, e.g., regulators, consumers, or competitors. This perspective is commonly referred to as the ``proprietary cost" hypothesis. The second approach, introduced in models with uncertain information endowments \'{a} la \citeasnoun{dye85} and \citeasnoun{junkwo88}, assumes that being genuinely uninformed is non-verifiable. In these settings, an informed party may pretend to be uninformed, leveraging the resulting ambiguity to strategically withhold unfavorable information. 

In this study, we demonstrate that these approaches are special cases of a unified theory, framed in terms of a privately-observed random disclosure cost. Specifically, when the distribution of the disclosure cost is concentrated at a single value (a mass point), the theory simplifies to a fixed disclosure cost model. Conversely, if the cost is randomly either zero or infinite, the model corresponds to uncertain information endowment. More generally, outside of these special cases, the model accommodates more general patterns of cost uncertainty, including dependence on privately-observed information in environments where costs are potentially correlated to fundamentals.

By formalizing the theory, we demonstrate that seemingly distinct approaches belong to a single theory,
whose many specifications share common intuitions.
That is, the researcher can use the theory of random costs to pursue general insights without binding the argument to assumed, but empirically unobservable details about the friction. We further show  that the theory nests asset sales under asymmetric information \cite{ake70,lev01}. In this setting, the manager chooses whether to retain the asset, incurring a liquidity cost, representable as a ``disclosure cost," or sell the asset at an uninformed  ``non-disclosure" price.
We show, more generally, that the theory can be formulated in terms of a cost distribution that is correlated with privately-held information, thus accommodating, without changes to the formalism, plausible applications in which costs are correlated with fundamentals \cite{fri24}. 

In pursuing a generalized disclosure theory, we want the exercise to offer more than a concatenation of the multiple models, in the sense that the whole should be more than the sum of its parts. The framework uses single unifying concept to combine several theories, rather than stating each theory as separate economic forces. For example, alternatively, it is readily verified that equation (7) in \citeasnoun{junkwo88} can be adjusted if disclosure features a fixed cost $c>0$ to yield
\begin{equation}\label{FRI}
q(\mathbb{E}(\tilde{x})-\tau+c)=(1-q)\int_{-\infty}^{\tau}F(x)dx-(1-q)F(\tau)c,
\end{equation}
where $\tau$ is the disclosure threshold, $q$ is the probability of being uninformed, and $F(.)$ is the distribution of the firm value $\tilde{x}$.\ Equation (\ref{FRI}) separately incorporates two frictions but does not reconcile them into a single economic force.
 Further, the goal of the unification of the theory is to allow for the generation of new theories that could be fundamentally different from the special cases. In particular, equation (\ref{FRI}) does not easily accommodate costs that are random or depend on $\tilde{x}$, because such an assumption will change the non-disclosure expectation and affect the right-hand side integral; similarly, the connection between disclosure and the market for lemons is not visible from an approach that combines fixed costs and uncertainty about information endowment.

We discuss the main results below. We establish that the equilibrium, which is usually stated in terms of a threshold type for the marginal discloser, can be 
re-expressed in terms of a straightforward equation that determines the non-disclosure price. Unlike a traditional threshold, the characterization holds more generally for  costs that are stochastic and may be correlated with the underlying signal, providing a parsimonious analysis of optimal disclosure policies across a wide class of environments.
Using this expression, we show that an equilibrium always exists. Even if there is a full-disclosure (e.g., unraveling) equilibrium, under mild conditions, there exists at least one equilibrium featuring strategic withholding.

We also demonstrate that full-disclosure equilibria exist under broader conditions than typically assumed in prior literature, as long as  verification costs do not exceed total value. We argue that this is a reasonable presumption to have on any communication, as noted in the argument below.\ Consider the extreme case where there are some (largely unverifiable) messages whose disclosure costs exceed total firm value: for example, messages whose verification cost would be infinite.\ Normalizing the minimum value to zero, suppose that firms can publicly incur a cost \textit{lower} than this verification cost but, if they do so, their disclosure is unverifiable or cheap talk.\footnote{In practice, one would envision a game where firms have access to different degrees of verifiability, increasing in the cost they are willing to spend. The argument stated here applies to information becoming unverifiable below a given cost, so it would naturally apply to any such environment as well.} Then, there is always a fully-revealing equilibrium in which all firms incur the verification cost
if the cost is less than their value and, otherwise, spend a disclosure cost equal to their value leading to an unverified disclosure.
This is an equilibrium because
all non-verified disclosures achieve a surplus\ (i.e., their price net of cost) of zero
versus a negative surplus if they paid the verification costs, barring any profitable deviation; vice-versa, firms choosing a verified disclosure achieve a positive surplus and are better-off than if they choose any unverifiable message (at lower cost). Surprisingly, and in admittedly shocking contradiction to \citeasnoun{ver83} and \citeasnoun{dye85}, we derive unraveling equilibria in these settings by adding costly unverified communication but no change to the arrival of information
or the verification technology. Formalizing the argument requires a generalized perspective on disclosure costs that is simply not apparent in standard models.

The existence of multiple equilibria  poses challenges to the empirical content of the theory. In particular, otherwise reasonable environments feature equilibria with counter-intuitive comparative statics, such as disclosure \textit{increasing} in disclosure costs. To address the multiplicity, we introduce a refinement based on neologism-proofness, a forward-induction argument originally proposed in the context of cheap talk \cite{far93} and recently adapted in disclosure games \cite{bercia18}. This refinement assumes that the manager can propose a rational interpretation of the non-disclosure. The resulting equilibrium features the maximal non-disclosure price and, therefore, full unraveling only occurs if it is the unique equilibrium outcome. As a counter-point to the minimum principle \cite{achdemkre11,gutkreskr14}, which establishes that without costs, an equilibrium minimizes the non-disclosure price, the ``maximum" principle  establishes that the non-disclosure price is maximized among all equilibria and holds among more general classes of games with (potentially)\ non-zero costs.

We analyze next the case in which disclosure costs are type-independent but random, deriving a clean expression for the equilibrium non-disclosure price using a convolution integral over the cost and signal distributions. The partial disclosure equilibrium is unique under a monotone likelihood ratio assumption and exists whenever the signal distribution has a thin lower tail. Conversely, for distributions with heavy tails indicative of a high frequency of large adverse outcomes,  unraveling may persist even in the presence of non-zero fixed costs, as extreme losses induce disclosure despite moderate frictions. The finding disproves the common, but incorrect, statement that disclosure costs ``imply" strategic non-disclosure.\footnote{For example, \citeasnoun{beycohlys10} note ``In the absence of any disclosure, investors rationally infer that the
value of the firm's assets is low and/or the underlying risk is high. Despite the fact that investors
view the absence of disclosure as relatively ``bad news," the manager does not disclose
sufficiently unfavorable news because he can achieve a higher market price by avoiding the costs
associated with a disclosure." }

The model is applied to empirical settings by showing that the disclosure decision maps naturally into a standard discrete choice empirical design. The binary decision to disclose or withhold can be interpreted as the outcome of a latent variable model in which the difference between expected firm value and disclosure cost determines the manager's action. This structure implies that the probability of disclosure follows a probit model with parameters that can be structurally identified from the data. We estimate the model using a sample of management earnings forecasts and show that the data strongly support the presence of random disclosure frictions. Notably, the distribution of forecasted news includes a substantial mass of negative surprises, which would be ruled out under classical disclosure models without random costs. Our estimated structural parameters indicate both a high average cost and significant cost dispersion, validating the theoretical assumption that disclosure costs are privately observed and heterogeneous.

Our objective of nesting several canonical disclosure environments within a single framework is related to the mechanism-design agenda pursued in \citeasnoun{bonhem21} and \citeasnoun{bonhem25}, who propose a more general unification in which disclosure incentives are derived endogenously from an optimal contracting problem. In their approach, equilibrium partial communication (and, in particular, upwardly biased voluntary disclosures) can arise even without positing exogenous disclosure costs, because limited commitment and renegotiation-proofness restrict the set of implementable truth-telling mechanisms. Their model also encompasses types of communication that our model does not capture, such as possibly unverifiable supplementary disclosures \cite{lankri08,bermar16,ein18}.
By contrast, we take the evidence-game stage as primitive and provide a reduced-form unification across models in which disclosures are verifiable when made, but may be strategically withheld; all such frictions are summarized by a privately observed random disclosure cost (or benefit) whose distribution may depend on fundamentals. The key distinction is thus one of \textit{where} the friction enters: \citeasnoun{bonhem25} endogenize the wedge through optimal mechanism design and  commitment constraints, whereas we parameterize the wedge directly in the sender's disclosure payoff. This difference also makes the approaches complementary: their framework provides micro foundations for economically meaningful distributions of disclosure incentives, while ours yields a tractable equilibrium characterization and comparative statics that apply across a class of evidence games and map naturally to empirical discrete-choice specifications.

Finally, there is also a well-established literature on evidence games that is closely related to our objective. This literature identifies a variety of environments in which the canonical prediction that firms disclose only favorable information need not obtain (see, e.g., \citeasnoun{kir97}, \citeasnoun{levsmi03}, \citeasnoun{ein05}, \citeasnoun{pae05}, \citeasnoun{sui07}, \citeasnoun{beygut12}, and \citeasnoun{lansiv10}). Our model contributes to this literature by showing how uncertainty about disclosure costs, rather than their level alone, can generate systematic disclosure of unfavorable news.

Section \ref{sec:model} presents the general framework and main assumptions. Section \ref{section3} presents necessary and sufficient conditions for unraveling, and examines conditions for existence, selection, and uniqueness of equilibria. Section \ref{part2} focuses on the special case where disclosure costs are random but independent of fundamentals to further examine the impact of the cost distribution on the probability of disclosure. Section \ref{estimation} estimates the structural parameters of the model using a discrete-choice (probit) approach. Section \ref{conclusion} concludes.

\section{A Unified Disclosure Theory} \label{sec:model}

%\subsection{Assumptions and\ Equilibrium}
\paragraph{}
Below, we present a general theory in which a unified friction is modeled as the sender's privately observed utility or disutility from achieving an information-sensitive payoff. Depending on specifications of this utility, the theory yields as special cases several classic models: the unraveling theorems absent frictions \cite{vis78,mil81,gro81}, the uncertain information endowment model of \citeasnoun{dye85} and \citeasnoun{junkwo88}, the fixed cost model of \citeasnoun{jov82} and \citeasnoun{ver83}, the market for lemons of \citeasnoun{ake70}. In addition, the model has many other applications, such as contexts in which these frictions jointly occur, such as disclosure costs followed by trading under asymmetric information \cite{bervayxue21}, mandatory disclosure \cite{aghcorzhe21}, generalized disclosure costs \cite{sui05,cheziv21,fri24}, voluntary disclosure in mechanism design \cite{bonhem25}, or structural estimation applications \cite{bermamar20,cheliu20,bermarter22}, where disclosures may be inconsistent with strict sanitization \cite{shi94}.

Hereafter, for expositional purposes, we use the context of  a ``manager" disclosing information to ``investors" in a competitive capital market subject to a random cost or benefit, e.g., a forecast or financial statement containing ex-post verifiable facts about the firm. However, the analysis can be interpreted more generally to sender-receiver games in which a sender informed about a true state discloses selectively to increase a receiver's posterior expectation, and the receiver sets an action to match the true state; see \citeasnoun{sto13} for discussions of this class of models.\footnote{These models are often referred to as ``evidence games" because the sender can present or withhold privately-observed evidence, see, for recent examples, \citeasnoun{ein07}, \citeasnoun{einziv08}, \citeasnoun{gutkreskr14}, \citeasnoun{harkreper17}, \citeasnoun{glooppzha18}, \citeasnoun{frihugmic22}, \citeasnoun{licwek23}, \citeasnoun{farfreisp24}, \citeasnoun{per24}, and \citeasnoun{kimpae25}. With a few exceptions, most evidence games do not allow for message-specific costs; by contrast, our approach shows that a generalized formulation with random privately-observed costs nests as a special case the random receipt of evidence.}
  
Formally, the manager has verifiable information, denoted as a random variable $\tilde{x}$ indicating their belief about the expected value of the firm. It is common-knowledge that $\tilde{x}$ has a  p.d.f. $f(.)$, c.d.f. $F(.)$, finite mean $\mu$, and full support over $\mathbb{R}^+$. To reduce notations, the manager is assumed to be risk-neutral, as all proofs are unchanged if the manager is risk-averse as long as we rewrite all payoffs in  utility space. We set the support of $\tilde{x}$ to be bounded from below: this assumption seemingly rules out common distributions such as the normal but is not critical for most results. Relaxing it will require technical considerations over the lower tail to be made explicit in Section \ref{part2}.

We depart from standard disclosure models by assuming that the disclosure cost $\tilde{c}$ is a random variable. While known to the manager before disclosing, it is unobserved by outside parties and cannot be separately disclosed. The manager makes a binary decision whether to disclose or withhold their entire private information, so that either investors (i) observe the realized $\tilde{x}-\tilde{c}$, or (ii) do not observe any information. To be more precise, the manager cannot selectively withhold $\tilde{x}$ while disclosing the (counter-factual) disclosure cost $\tilde{c}$, and does not experience additional cost from disclosing $\tilde{c}$ when $\tilde{x}$ is disclosed. This implies that, even if $\tilde{c}$ could be withheld after a disclosure of $\tilde{x}$, standard unraveling arguments would imply that it would always be disclosed voluntarily \cite{mil81,gro81}.\footnote{We may also assume, with the same implications on the model solution, that the cost cannot be disclosed because it is privately borne by the manager (e.g., a cost of time and effort to make information verifiable).}

In what follows, the notation $(\tilde{x},\tilde{c})$ is used for random variables with, in short-hand, $(x,c)$ for realizations. The disclosure cost has a c.d.f. $H(.|x)$, which may also capture contexts in which the value $\tilde{x}$ and the disclosure cost $\tilde{c}$ are correlated. We possibly allow $H(.|x)$ to be a mass point, in which case the cost may be deterministic. Further, this distribution may also have support over negative values if, for example, the firm might face a penalty from non-disclosure or some benefits from coordinating with its competitors \cite{boushezal20}. 

To draw a parallel to models incorporating uncertain information endowment \cite{dye85,junkwo88}, the concept of cost extends beyond monetary expenses incurred by the firm upon disclosure. Instead, this cost represents a broader relative disutility, encompassing various impediments or benefits associated with disclosure. Examples include the verification of information, potential litigation risks arising from either disclosure or nondisclosure (e.g., from employees, regulators, or competitors), expenses related to collecting and organizing data, and situations where the firm or investors fail to obtain information (i.e., $c=\infty$).
Specifically, this characterization is consistent with a revealed preference representation, where any friction influencing a manager's inclination to disclose information is modeled as a privately-observed random variable incorporated directly into the manager's payoff structure. 

As noted in \citeasnoun{fri24}, the mechanisms described by the proprietary costs hypothesis suggest that it is unlikely for  $\tilde{c}$   to possess a single mass point independent of $\tilde{x}.$ Furthermore, we argue that this implies that $H(.|x)$ is not a deterministic function of $\tilde{x}$, given that proprietary costs may depend on privately observed information not perfectly correlated with firm value.
Certain firms may experience disclosure-related costs due to competitive pressures, making these costs a function of other privately observed factors such as whether firms' customers are captive or protected from competitors, or whether the firm's innovations and technologies can easily be replicated. Conversely, disclosure may also yield benefits, such as signaling a strong competitive advantage and deterring competition. In the end, the firm is likely better informed about the precise implications of disclosure than external observers, aligning with our assumption that disclosure costs are privately known.

After privately observing $(x,c)$, the manager decides whether to disclose or withhold in a competitive market, in the form of a report $r(x,c)\in\{x-c,\emptyset\}$. The market forms rational expectations, setting $P(x-c)=x-c$ conditional on disclosure and 
\begin{equation}\label{pempty}
P(\emptyset)=\mathbb{E}(\tilde{x}|r(\tilde{x},\tilde{c})=\emptyset),
\end{equation}
conditional on non-disclosure. 

The manager follows a strategy to maximize the current price
such that, without loss of generality, disclosure is chosen when indifferent between disclosure and non-disclosure. This implies that, for an equilibrium non-disclosure price $P(\emptyset)\in \mathbb{R}^+$, an optimal strategy $r(x,c)$ must satisfy $r(x,c)=x-c$ if and only if $P(x-c)=x-c\geq P(\emptyset)$. It is then readily verified that the disclosure strategy can be written as a threshold strategy $\tau(c)\equiv c+P(\emptyset)$ such that the manager discloses if and only if $x\geq \tau(c)$. Naturally, if $\tilde{c}$ is non-random, the problem reduces to a  threshold $\tau$ taking the constant cost as a  parameter.

The standard approach is to characterize the threshold $\tau(c)$, using it to substitute out $P(\emptyset)$ from $(\ref{pempty})$ in the indifference condition. However, doing so is impractical here because the threshold $\tau(c)$ is a frontier in the space $(x,c)$. We adopt instead an equivalent formulation by characterizing the (scalar) non-disclosure price $P(\emptyset)$ so that, after substituting $\tau(c)$ in $(\ref{pempty})$, one is left with one equation in one unknown.  Specifically, the manager discloses  if $x-c\geq P(\emptyset)$ which, from (\ref{pempty}), implies the following expectation:
\begin{eqnarray}\label{xx1}
P(\emptyset)&=&\mathbb{E}(\tilde{x}|\tilde{x} \ < \ P(\emptyset)+\tilde{c}).
\end{eqnarray}
In cases where the manager is indifferent between disclosure and non-disclosure, we assume that disclosure is selected. 
\begin{definition1}\label{J1}
An equilibrium is defined as a non-disclosure price $P(\emptyset)$ and a disclosure threshold $\tau(c)=P(\emptyset)+c$, such that the firm discloses if and only if $x\geq \tau(c)$ and $P(\emptyset)$ satisfies (\ref{xx1}) when possible. An unraveling equilibrium is such that $r(\tilde{x},\tilde{c})=\tilde{x}-\tilde{c}$ almost surely.
\end{definition1}

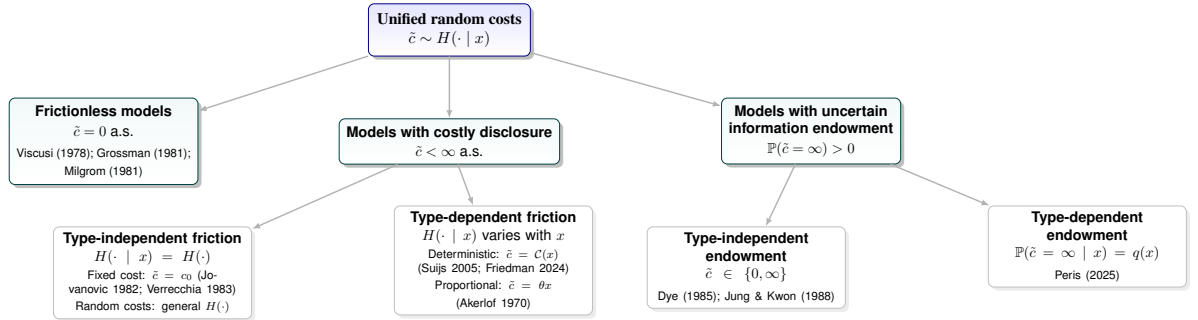
\begin{figure}[h]
\centering
\resizebox{0.98\textwidth}{!}{
\begin{tikzpicture}[
    font=\small\sffamily,
    node distance=10mm and 16mm,
    % Root Node Style
    root box/.style={
        draw=blue!50!black, 
        top color=blue!10, 
        bottom color=white, 
        rounded corners, 
        align=center, 
        inner sep=6pt, 
        drop shadow, 
        font=\bfseries
    },
    % Level 1 Node Style
    level1 box/.style={
        draw=teal!50!black, 
        top color=teal!5, 
        bottom color=white, 
        rounded corners, 
        align=center, 
        inner sep=5pt, 
        drop shadow
    },
    % Level 2 Node Style
    level2 box/.style={
        rectangle,
        draw=gray!50, 
        fill=white, 
        rounded corners, 
        align=center, 
        inner sep=4pt,
        text width=4.5cm, % Ensures consistent width for leaves
        drop shadow={opacity=0.15}
    },
    % Connector Lines
    edge/.style={
        draw=gray!60, 
        line width=1pt, 
        -{Latex[length=2mm]}
    }
]

% Root
\node[root box] (root) {Unified random costs\\$\tilde c \sim H(\cdot\mid x)$};

% Level 1
\node[level1 box, below left=of root, xshift=-25mm] (fric)
    {\textbf{Frictionless models}\\[0.5mm] $\tilde c=0$ a.s.\\[1mm] \scriptsize Viscusi (1978); Grossman (1981);\\ \scriptsize Milgrom (1981)};

\node[level1 box, below=of root, yshift=-5mm] (cost)
    {\textbf{Models with costly disclosure}\\[0.5mm] $\tilde c<\infty$ a.s.};

\node[level1 box, below right=of root, xshift=30mm] (endow)
    {\textbf{Models with uncertain}\\\textbf{information endowment}\\[0.5mm] $\mathbb{P}(\tilde c=\infty)>0$};

% Level 2
% Note: Kept original positioning logic provided by user

% Under Cost
\node[level2 box, below left=of cost, xshift=-5mm, yshift=-5mm] (costTI)
    {\textbf{Type-independent friction}\\ $H(\cdot\mid x)=H(\cdot)$\\[1mm] \scriptsize Fixed cost: $\tilde c=c_0$ (Jovanovic 1982; Verrecchia 1983)\\ \scriptsize Random costs: general $H(\cdot)$};

% Under Endow (but visually placed below left of Endow)
\node[level2 box, below left=of endow, xshift=-15mm] (costTD)
    {\textbf{Type-dependent friction}\\ $H(\cdot\mid x)$ varies with $x$\\[1mm] \scriptsize Deterministic: $\tilde c=\mathcal{C}(x)$ (Suijs 2005; Friedman 2024)\\ \scriptsize Proportional: $\tilde c=\theta x$ (Akerlof 1970)};

% Under Cost (but visually placed below right of Cost)
\node[level2 box, below right=of cost, xshift=5mm, yshift=-5mm] (endTI)
    {\textbf{Type-independent endowment}\\ $\tilde c\in\{0,\infty\}$\\[1mm] \scriptsize Dye (1985); Jung \& Kwon (1988)};

% Under Endow
\node[level2 box, below right=of endow, xshift=5mm] (endTD)
    {\textbf{Type-dependent endowment}\\ $\mathbb{P}(\tilde c=\infty\mid x)=q(x)$\\[1mm] \scriptsize Peris (2025)};

% Edges
\draw[edge] (root) -- (fric);
\draw[edge] (root) -- (cost);
\draw[edge] (root) -- (endow);

\draw[edge] (cost) -- (costTI);
\draw[edge] (cost) -- (costTD); % Preserved original connection logic

\draw[edge] (endow) -- (endTI);
\draw[edge] (endow) -- (endTD);

\end{tikzpicture}
}
\caption{Hierarchical mapping of classical voluntary disclosure models nested within the unified framework.}
\label{LH}
\end{figure}

%\vspace{.5cm}

%\vspace{.5cm}
In Figure \ref{LH}, we illustrate that, depending on the choice of the distribution $H(c|x)$, the model branches into several common sub-theories, depending on the support of $\tilde{c}$. On the first branch, models in which $\tilde{c}$ is zero with probability one collapse to unraveling theory, such that all managers disclose with probability one. On the second branch, models in which $\tilde{c}|\left(\tilde{x}=x \right)$ is non-random but potentially positive, separate further into models in which the cost is fixed, or is increasing in the information received, further separating into information-contingent proprietary cost or the market for lemons. On the third branch, if $\tilde{c}$ is random, the model specializes to type-contingent uncertainty about information endowment, and as a special case, random arrival of information. Each of these models is mapped to a disclosure cost in more detail below. 

\vspace{.5cm}
\paragraph{}
\textit{Example 1: Unraveling theory \cite{vis78,gro81,mil81}.} If $H(c|x)=1_{c\geq 0}$, the firm faces no friction because $\tilde{c}=0$ almost surely.
\paragraph{}
\textit{Example 2: Disclosure cost \cite{jov82,ver83}.} If $H(c|x)=1_{c\geq c_0}$, the firm incurs a fixed disclosure cost, independent of the private information.
\paragraph{}
\textit{Example 3: Uncertain information endowment \cite{dye85,junkwo88}.} If $H(c|x)=q \times 1_{c\geq 0}$, the model corresponds to a probability $q\in (0,1)$ of receiving information, mapping the state in which information is received to $\tilde{c}=0$ and the state in which information is not received to $\tilde{c}=\infty$ (i.e., such that it would be infinitely costly to disclose it for any $x$).
\paragraph{}
\textit{Example 4: Contingent costs \cite{sui05,bercia18,fri24}.} If $H(c|x)=1_{c\geq \mathcal{C}(x)}$, the model corresponds to a deterministic cost function $\mathcal{C}(x)$, which may depend on the information received.
\paragraph{}
\textit{Example 5: Type-dependent endowment \cite{per24}.} Example 3 can be further generalized to $H(c|x)=q(x) \times 1_{c\geq 0}$, in which case the probability of receiving information may depend on the value of the firm.
\paragraph{}
\textit{Example 6: Market for lemons \cite{ake70}.} If $H(c|x)=1_{c\geq \theta x }$ (and $\tilde{x}\sim U[0,1]$), the model reduces to the market for lemons, after re-interpreting the disclosure payoff as the value from retaining the asset $(1-\theta)x\in (0,x)$ and the non-disclosure price as the value from selling the asset $P(\emptyset)=\mathbb{E}(\tilde{x}|(1-\theta)\tilde{x}\leq P(\emptyset))$. Note that this is a special case of example 4 with a cost proportional to $x$.
\paragraph{}
\textit{Example 7: Mandatory Disclosure \cite{mar13,guaver18,aghcorzhe21}.} If the cost may be $c=-\infty$ with positive probability, that is $\lim_{c\rightarrow -\infty} H(c|x)= q(x)$, the model can accommodate a positive probability of mandatory disclosure.

\paragraph{}
Each of these examples reflects different mechanisms but shares a common feature. The manager makes a decision between conditioning their payoff on their private information, or achieving a payoff that is not (directly) type-contingent. Example 1 is the classic model in which costs are omitted from the model. In examples 2-5 and 7, there is a disclosure decision that reveals the payoff prior to a sale. In example 6, the manager realizes the payoff by holding the asset, or can resell the asset, in which case no information directly realizes in payoffs except what is conveyed by the selling decision.

\section{Equilibrium: Existence, Uniqueness, and Selection}
\label{section3}
\paragraph{}
We first examine conditions under which unraveling occurs. It can be readily verified that, if it exists, the unraveling equilibrium may always be sustained with the most pessimistic non-disclosure belief $P(\emptyset)=0$. The next Proposition develops a simple necessary and sufficient condition for unraveling to hold. Except for arguments conveying the main economic intuition presented in the main text, all proofs are relegated to the Appendix.

\begin{proposition1}\label{unr}
There exists an unraveling equilibrium if and only if, almost surely, $\tilde{x}-\tilde{c}\geq 0$.
\end{proposition1}

The existence of disclosure costs is not sufficient to rule out unraveling: disclosure costs must be sufficiently large to offset the benefits of disclosure. Indeed, Proposition \ref{unr} formalizes this claim in terms of a relatively mild condition: an unraveling equilibrium exists as long as the cost is never greater than the value implied by the worst information. Stated in converse form, for unraveling to fail, disclosure must be economically infeasible for certain managers because it would imply a price (net of cost) strictly worse than costlessly revealing the worst information. 

Interestingly, this condition will always be met as long as one allows for a seemingly innocuous extension of the message space,  even without changing the verification technology. Suppose that a manager observing a realization $(\tilde{x},\tilde{c})=(x,c)$ can publicly pay a cost $y\in \mathbb{R}^+$, which may be strictly lower than the required disclosure cost $c$. To avoid altering the verification technology, assume that, when $y<c$, the resulting message is unverifiable, serving no informational purpose; for example, we may interpret this as a low-quality verification or a costly but wasteful accounting system.\footnote{We intentionally make the extreme assumption that a verification with $y<c$ leads to pure cheap talk to build an argument for unraveling that does not involve additional verification. Of course, in practice, we might assume that there is some degree of (partial) verification for $y<c$ (say, if the signal $x$ were revealed with some probability decreasing in $c-y$), which implies the same insight.} Now consider a candidate equilibrium in which the manager chooses $y=\min(c,x)$, which is fully-revealing because either $y=x<c$ reveals the type from $y$ or $y=c$ yields a verified message. In this case, all \textit{unverified} reports achieve a surplus of zero (since they are priced at $x$ and pay $y=x$) versus a negative surplus $x-c<0$ if they were verified. All \textit{verified} reports achieve $x-c>0$, which is better than any zero-surplus unverified report. Hence, these strategies and prices are indeed an equilibrium. In this equilibrium, disclosure is more frequent than suggested in the threshold equilibrium of \citeasnoun{jov82} and \citeasnoun{ver83}.

The equilibrium construction applies to models featuring uncertain information endowment. Suppose that all informed managers disclose their information while the uninformed manager sets $y=\mu$; the market price is then set to zero for any non-disclosure with $y< \mu$ and equal to $\mu$ if $y\geq \mu$. First, informed managers with $x\in (0,\mu)$ are obviously worse off if they choose $y\geq \mu$ or if they choose $y<\mu$, since in the latter case they achieve a market price of zero.  Second, informed managers with $x\geq \mu$ are weakly worse-off if they do not disclose, given that the highest market belief is $\mu<x$. Third, uninformed managers achieve a net price of zero in equilibrium, which is weakly greater than any other $y'<\mu$ given that these other choices yield a price of zero. As in the previous example, the equilibrium similarly implies more disclosure than in \citeasnoun{dye85} and \citeasnoun{junkwo88}, as there is no longer any strategic withholding.

We formally prove the above argument in the general model by considering the disclosure game of Section \ref{sec:model}, augmented as follows. After observing $(x,c)$, the manager can publicly choose a cost $y\in\mathbb{R}_+$. If $y\ge c$, verification succeeds and investors observe $x$. If $y<c$, verification about $x$ fails and investors observe only the expenditure $y$. Similarly, the manager's payoff is the market price minus the cost $y$. Proposition \ref{prop:money_burning} shows that a fully revealing equilibrium always exists in this extended game with costly unverified communication.

\begin{proposition1} \label{prop:money_burning}
In the game with an extended message space, there exists a perfect Bayesian equilibrium in which the market price equals $x$ almost surely.
\end{proposition1}

While unraveling cannot be ruled out, there may be other partial disclosure equilibria. Even if we restrict to $\tilde{c}\leq \tilde{x}$ (i.e., implicitly allow for an extended message space), the resulting costs to reveal the information could be very large, thus casting some doubts as to whether the manager would naturally coordinate on such an outcome. To illustrate this further, in the first case of fixed disclosure costs, the entire surplus of all firms who cannot profitably disclose is zero; in the second example of uncertain information endowment, uninformed firms similarly achieve a surplus of zero. In other words, non-disclosure equilibria may still offer a more ex-ante desirable outcome for the manager. In what follows, we develop this intuition in more detail: first, showing conditions for the existence of a partial disclosure equilibrium and, second, providing a neologism-based refinement demonstrating when such an equilibrium would plausibly occur.

Excluding the case of unraveling already discussed in Proposition \ref{unr}, non-disclosure has positive probability and thus one can write the right-hand side of (\ref{xx1}) as
\begin{eqnarray} %\label{c}
P(\emptyset)&=&\frac{\int x\overline{H}(x - P(\emptyset)|x)dF(x) }{\int \overline{H}(x - P(\emptyset)|x)dF(x)}, \label{xx}
\end{eqnarray}
where $\overline{H}(c|x)\equiv 1-H(c|x)$ and the denominator represents the probability of non-disclosure while the numerator integrates the possible firm values $\tilde{x}$ given its distribution conditional on non-disclosure. We define a partial disclosure equilibrium as an equilibrium such that the denominator of this equation is non-zero.

\begin{proposition1}\label{exist}
If there is no unraveling equilibrium, there exists at least one partial disclosure equilibrium. By contrast, suppose an unraveling equilibrium exists, i.e., $H(x| x)=1$ for all $x$. Assume further that for each $x$, $H(\cdot| x)$ has full support on $[0,x]$ with continuously differentiable density $h(\cdot| x)$, and satisfies  $\lim_{c\uparrow x} \frac{\partial}{\partial c}h(c| x)<h(x|x)=0$. Then at least one partial disclosure equilibrium exists.
\end{proposition1}

Proposition \ref{exist} establishes a set of sufficient conditions such that a disclosure equilibrium exists even if unraveling is possible. The first condition $H(x|x)=1-h(x|x)=1$ is relatively mild and is primarily made to rule out knife-edge cases where costs are known with certainty or some potential costs are so large, i.e., above or equal to firm value, that unraveling equilibria become implausible. The second part of the condition  $\lim_{c\uparrow x} \frac{\partial}{\partial c}h(c| x)<0$ is more demanding and requires a mass of ``high costs" firms. This condition generalizes the assumption, made in standard models, that verification costs are strictly higher than firm value; indeed, we formally show that one needs only to assume that some costs are potentially close enough to firm value.\footnote{It is not critical for this argument to assume that $\lim_{c\uparrow x} \frac{\partial}{\partial c}h(c| x)$ is non-zero. If this condition is not satisfied, the proof extends to a more general condition such that there exists $n\geq 1$ such that $\lim_{c\uparrow x} \frac{\partial^k}{\partial c^k}h(c| x)=0$ for any derivative $k<n$ and $\lim_{c\uparrow x} \frac{\partial^n}{\partial c^n}h(c| x)$. In other words,  it is sufficient for the function $h(x|x)$ to be analytic (i.e., admit a Taylor expansion).}

While Proposition \ref{exist} shows that at least one equilibrium exists, the model typically has multiple equilibria. When unraveling is possible, it is unclear whether markets would coordinate on an equilibrium featuring less informative prices. Further, even among partial disclosure equilibria, some of these equilibria further exhibit atypical comparative statics that seem to rule out even basic comparative statics, unless one places (somewhat ad hoc) technical restrictions on the distributions $H(.|x)$ and $F(x)$. 

To illustrate this problem, consider a fixed cost $\tilde{c}=3$ and suppose that $\tilde{x}=e^{\tilde{y}}$ is a lognormal mixture with $\tilde{y}$ drawn from  $N(1,1/3)$ with probability $1/3$ and $N(2,1/10)$ with probability $2/3$.
Note that while the resulting distribution is not logconcave \cite{bagber05} or (even) single-peaked, there are a priori no economic reasons to presuppose that distributions should be logconcave, except for mathematical convenience.
For example, this model may be interpreted as uncertainty about the mean return, which may be drawn from two possible distributions. 
\begin{figure}
\begin{center}\includegraphics[scale=.8]{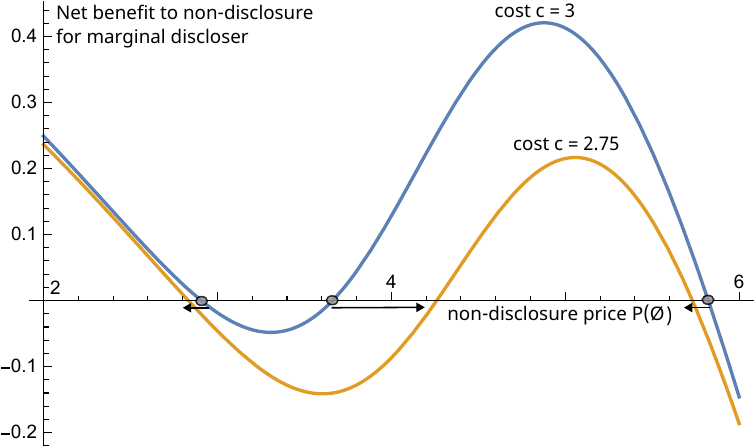}
\caption{Multiple Nash\ equilibria (gray circle)\label{L1}}\end{center}
\end{figure}

In  Figure \ref{L1}, we plot the function $ \psi(P(\emptyset))\equiv P(\emptyset)-\mathbb{E}(\tilde{x}|\tilde{x} \, < \, P(\emptyset)+\tilde{c})$ as a function of $P(\emptyset)$ whose zeros indicate Nash equilibria. There are, in this example, \textit{three} Nash equilibria.\ The left-most zero, at $P(\emptyset)\approx 2.9$ has the lowest non-disclosure price, corresponding to the equilibrium that is suggested by the draw from $N(1,1/3)$. This equilibrium non-disclosure price largely ignores draws from the higher distribution $N(2,1/10)$  because such draws are unlikely conditional on $\tilde{x}-\tilde{c} \, < \, 2.9$. At the other extreme, the right-most zero, at  $P(\emptyset)\approx 5.8$ has the highest non-disclosure price corresponding to the equilibrium that is suggested by both draws. There is also an interior equilibrium at $P(\emptyset)\approx 3.7$ such that additional draws from $N(2,1/10)$ balance the lower draws from $N(1,1/3)$, leading to an equilibrium between the two extremum equilibria.

While it may seem appealing to pick the interior equilibrium given that it  ``averages" across equilibria and may seem a safer robust approximation when it is unclear which equilibrium will be played, this equilibrium presents a problem. We plot $\psi(P(\emptyset))$ after decreasing the cost to $\tilde{c}=2.75$, which should presumably increase disclosure or, equivalently, decrease the non-disclosure price. But as shown by arrows, this is not the case: while the two extremum equilibria exhibit less disclosure, as expected,  the interior equilibrium counter-intuitively features \textit{more} disclosure in response to higher costs.

There are several possible approaches to address the multiplicity.\ A direct approach, discussed below, is to examine sufficient conditions such that there is a unique partial disclosure equilibrium. For example, if $c$ is constant and $F(.)$ is logconcave (i.e., $\frac{\partial \mathbb{E}(\tilde{x}|\tilde{x}<\tau)}{\partial \tau}<1$), the net payoff to disclosure relative to non-disclosure $\tau-c-\mathbb{E}(\tilde{x}|\tilde{x}<\tau)$ is decreasing.
Hence, the equilibrium disclosure threshold $\tau^*-c-\mathbb{E}(\tilde{x}|\tilde{x}<\tau^*)=0$, which makes the marginal firm indifferent, must be unique. However, this argument is limiting along several dimensions.\ First, logconcavity is a mathematical condition with no straightforward behavioral microfoundation; furthermore, it may or may not be satisfied in applications of the theory and, as shown in the example above, does not typically hold if there is heterogeneity in the population of interest. Therefore, we explore next a more general condition for uniqueness.

\begin{proposition1} \label{prop:unique}
Let $$\Lambda(y) \equiv \frac{\int x (1-H(x - y|x)) d F(x)}{\int (1-H(x - y|x)) d F(x)}.$$ 
The following holds: \\
(i) Suppose that $\tilde{x} - \tilde{c} \ge 0$ almost surely. Unraveling is the unique equilibrium if and only if $\Lambda(y) < y$ for all $y > 0$ such that $\Pr(\tilde{x} - \tilde{c} < y) > 0$;\\
(ii) Suppose that $\Pr(\tilde{x} - \tilde{c} \ge 0) < 1$. If $\sup_{y \in [0, \mu]} \frac{\int \left(x - \Lambda(y)\right) h(x - y|x) d F(x)}{\int \overline{H}(x - y|x) d F(x)} < 1$, there is a unique equilibrium and it features partial disclosure. Moreover, when $\tilde{x}$ and $\tilde{c}$ are independent, the partial disclosure equilibrium is unique if $\tilde{x}$ has a logconcave density, or $\sup_{x \ge c_0} c_0 \frac{f(x)}{F(x)} < 1$ where $\tilde{c} \equiv c_0$ is a fixed disclosure cost.
\end{proposition1}

In the subclass of models with uncertain information endowment $\tilde{c}\in \{0,\infty\}$, uniqueness can be directly verified without the condition from Proposition \ref{prop:unique}. Specific to these models, no firm incurs the disclosure cost in equilibrium so that the expected price across two tentative Nash equilibria must be equal to $\mu$. But, at the same time, each manager achieves $\max(P(\emptyset),\tilde{x}-\tilde{c})$ which implies that, for two candidate equilibria, $P_1(\emptyset)<P_2(\emptyset)$, the expected price of a manager before $(\tilde{x},\tilde{c})$ realizes is $$p P_1(\emptyset)+(1-p)\mathbb{E}(\max(P_1(\emptyset),\tilde{x})|\tilde{c}=0)<p P_2(\emptyset)+(1-p)\mathbb{E}(\max(P_2(\emptyset),\tilde{x})|\tilde{c}=0),$$
where $p\equiv \Pr(\tilde{c}=\infty)$.\ This contradicts that, by the law of iterated expectations, both sides must be equal to $\mu$. Note that this uniqueness result is shown without any algebraic manipulations, such as integration by parts \cite{junkwo88}, and holds more generally even if $\tilde{x}$ does not have a density or $\tilde{c}$ is correlated to $\tilde{x}$ \cite{per24}.

Since the first approach makes assumptions that may not always hold in practice, a second approach is to examine which equilibrium markets would plausibly coordinate on, using a refinement to narrow down the set of equilibria. %Below, we define such a criterion based on neologism-proofness \cite{far93}. Specifically, an equilibrium is neologism-proof if, given a particular Nash equilibrium, managers cannot propose an alternative message (a neologism) with a suggested interpretation that could be  tentatively assumed to be true and  credible to the managers who benefit from the interpretation. 
We adopt neologism-proofness \cite{far93} for this purpose. Specifically, an equilibrium is neologism-proof if, given a particular Nash equilibrium, managers cannot propose an alternative message (a neologism) with a suggested interpretation that could be  tentatively assumed to be true and  credible to the managers who benefit from the interpretation.

In the present disclosure environment, neologism-proofness is particularly simple because on equilibrium path disclosures are fully revealing; therefore, any neologism can be evaluated solely by its implications conditional on non-disclosure. Intuitively, if two partial disclosure equilibria exist with $P_1(\emptyset)<P_2(\emptyset)$, then the lower-price equilibrium is vulnerable to a credible neologism that proposes investors to interpret ``silence'' according to the belief conditional on non-disclosure in the higher equilibrium. Under standard forward induction logic (e.g., a small cost of sending the neologism), exactly those managers who benefit from the higher non-disclosure price would send it, making the proposed interpretation credible. We state the formal definition below.

\begin{definition1} \label{def:np}
Fix an equilibrium with non-disclosure price $P(\emptyset)$ and a disclosure rule $r(.)$ such that $r(x,c)=x-c$ if and only if $x-c\ge P(\emptyset)$. Add an off-equilibrium message $m$ (a ``neologism'') that does not disclose verifiable information. An interpretation of $m$ is a price $P(m)$ that investors set after $m$. The equilibrium is neologism-proof if there does not exist a neologism $m$ and an interpretation $P(m)$ such that: (i) the set of types that strictly benefit from $m$ at price $P(m)$ relative to the equilibrium is exactly the set of types that send $m$; and (ii) investors' price $P(m)$ equals the rational expectation of $\tilde x$ conditional on receiving $m$ given (i).
\end{definition1}

\begin{figure}[h!]
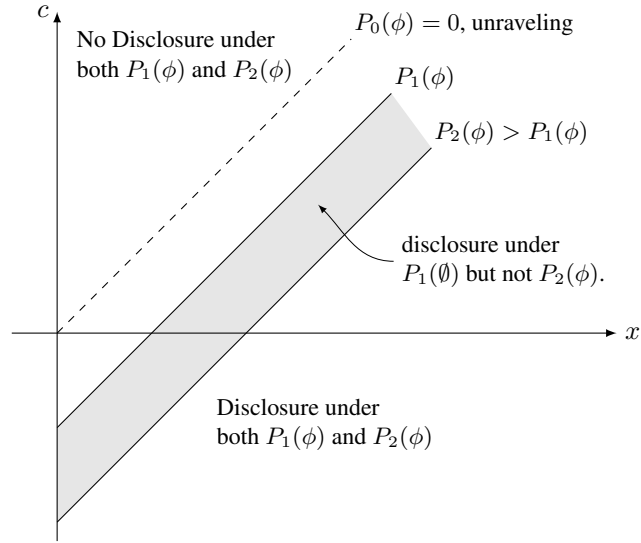

\begin{center}

\tikz[font=\small]{

\tikzset{every node/.style={inner sep=3pt}}

\draw[-latex] (.4,0)--node[pos=1,right] {$x$} +(8,0);
\draw[-latex] (1,-2.75)--node[pos=1,left] {$c$} +(0,7);

\draw[dashed] (1,0)--coordinate[pos=1] (a) node[pos=1, above right=-1mm, font=\footnotesize] {$P_0(\phi)=0$, unraveling} ++(45:5.5);
\draw (1,-1.25)--coordinate[pos=1] (b) node[pos=1, above right=-1mm] {\footnotesize $P_1(\phi)$} ++(45:6.25);
\draw (1,-2.5)--coordinate[pos=1] (c) node[pos=1, above right=-1mm] {\footnotesize $P_2(\phi)>P_1(\phi)$} ++(45:7);

%\draw[-latex] (.7,-.1)--(.7,-1.15);

\begin{scope}[on behind layer]
\fill[gray!20] (1,-2.5)--(c)--(b)--(1,-1.25)--cycle;
\end{scope}

\draw[latex-, very thin, font=\footnotesize] ($(c)+(-1.5,-.75)$) to[out=-45, in=180] node[pos=1, right, align=left] {disclosure under\\ $P_1(\emptyset)$ but not $P_2(\phi)$.} +(1,-.75);

\node[font=\footnotesize,align=left, anchor=south west] at (1.15,3.2) {No Disclosure under\\both $P_1(\phi)$ and $P_2(\phi)$};

\node[font=\footnotesize,align=left, anchor=north west] at (3,-.75) {Disclosure under\\both $P_1(\phi)$ and $P_2(\phi)$};
}
\end{center}
\caption{Comparing disclosure equilibria} \label{M1}
\end{figure}

To summarize, in the spirit of forward induction, the manager persuades the investor that such a neologism would have been senseless absent a belief that it would be taken seriously (by forward induction). In turn, this implies that all managers that benefit from the neologism would send it, thus challenging a tentative equilibrium. In this verbal argument, however, the last part of the neologism must be verified, namely, would it be the case that a subset of senders benefiting from the neologism would send it? To verify this argument, consider the three regions in Figure \ref{M1}.

\paragraph{}
 
\begin{enumerate}
\item[(i)] The north-west region, with $\tilde{x}-\tilde{c} < P_1(\emptyset)$ features non-disclosure under $P_1(\emptyset)$, so that managers in this region indeed prefer to send the neologism because it increases their payoff to $P_2(\emptyset)$.
\item[(ii)] The middle region, with $\tilde{x}-\tilde{c}\in [ P_1(\emptyset),P_2(\emptyset))$ features disclosure under $P_1(\emptyset) \le \tilde{x}-\tilde{c}$ but non-disclosure under $P_2(\emptyset)>\tilde{x}-\tilde{c}$. Thus, managers in this region are also better-off sending the neologism.
\item[(iii)] The south-east region, with $\tilde{x}-\tilde{c}\ge P_2(\emptyset)$, is such that senders would still disclose under the neologism and therefore have no incentives to send the neologism.
\end{enumerate}

\begin{proposition1}\label{exist0}
If there are multiple Nash equilibria $\mathcal{N}$, a neologism-proof equilibrium must maximize the non-disclosure price $P(\emptyset)$ in $\mathcal{N}$, i.e., features the lowest probability of disclosure.
\end{proposition1}

We conclude this section by noting that Proposition \ref{exist0} yields an extremum equilibrium, eliminating the equilibrium in Figure \ref{M1} that features increasing disclosure given higher costs. In fact, as $H(.|x)$ shifts in the sense of first-order stochastic dominance, the right-hand side of (\ref{xx1}) increases for any non-disclosure price, implying that there must exist an equilibrium with a higher non-disclosure price. This implies, as a result, that an increase in cost always decreases disclosure given a suitable neologism-proofness refinement. Furthermore, as stated next, unraveling equilibria must feature the lowest possible non-disclosure price; as a result, neologism-proofness always selects a partial disclosure equilibrium when one exists, thus providing a criterion to eliminate unraveling outcomes.\ 

\begin{corollary1}\label{exist012}
If there exists a partial disclosure equilibrium, unraveling is not neologism-proof.
\end{corollary1}

We conclude with two intuitive comparative statics. If $H(.|x)$ increases in the sense of first-order stochastic dominance, the right-hand side of (\ref{xx1}) must increase, since the non-disclosure price integrates over higher values of $\tilde{x}$. As a result, the maximal non-disclosure price that satisfies (\ref{xx1}) must increase when costs increase, thus demonstrating that greater frictions increase the non-disclosure price and thus decrease disclosure, as is intuitive.

A related comparative static can be obtained via certain scaling changes in uncertainty when $\tilde{c}\geq 0$. To show this, denote $\tilde{x}'=\alpha \tilde{x}$, with $\alpha>1$ indicating greater uncertainty and using $'$ to refer to all equilibrium objects suitably adapted to $\tilde{x}'$. Multiplying both sides of (\ref{xx1}) by $\alpha$ implies that the non-disclosure price under $\tilde{x}$ satisfies
\begin{equation*}
\alpha P(\emptyset)=\mathbb{E}(\alpha \tilde{x}|\alpha \tilde{x}< \alpha P(\emptyset)+\alpha \tilde{c}),
\end{equation*}
which implies that $\alpha P(\emptyset)$ is the equilibrium with maximal non-disclosure price for a model with cost $\alpha \tilde{c}>\tilde{c}$; it is also readily verified that this equilibrium must have the same probability of disclosure $\Pr(\alpha \tilde{x}-\alpha \tilde{c}\geq \alpha P(\emptyset))=\Pr( \tilde{x}-\tilde{c}\geq P(\emptyset))$ as $(\tilde{x},\tilde{c})$. However, because this model implies a distribution of costs $\alpha \tilde{c}$ that first-order stochastically dominates $\tilde{c}$, from our prior observation, $P(\emptyset)'<\alpha P(\emptyset)$ and features a higher probability of disclosure.\ 

To formalize these results, we consider two environments indexed by $\alpha \in\{\alpha,\alpha'\}$ where $H_\alpha(.| x)$ denotes the distribution of disclosure costs $\tilde{c}^\alpha$. Then, we say that the cost distribution under $\alpha'$ is \textit{higher} than under $\alpha$ if it satisfies first-order stochastic dominance and the monotone likelihood ratio property, that is, (i) $H_{\alpha'}(c| x)\le H_{\alpha}(c| x)$ and (ii) $\frac{1-H_{\alpha'}(x-y| x)}{1-H_{\alpha}(x-y| x)}$ is nondecreasing in $x$.

\begin{corollary1} \label{cor:cs}
Let $D_\alpha$ be the probability of disclosure in the equilibrium with maximal non-disclosure price: \\
(i) If the cost $\tilde{c}^{\alpha'}$ is higher than $\tilde{c}^\alpha$, then $D_{\alpha'} \le D_{\alpha}$. \\
(ii)  Suppose that $\tilde{x}_{\alpha}\equiv \alpha\tilde{x}$. If $\tilde{c}\ge 0$ almost surely and $\alpha' \tilde{c}$ is higher than $\alpha \tilde{c}$ (holding the distribution of $\tilde{x}_{\alpha'}$ fixed), then $D_{\alpha'}\ \ge\ D_{\alpha}$. 
\end{corollary1}

\section{Type Independent Costs}
\label{part2}

\paragraph{}
We move next to the case in which, while potentially random, $\tilde{c}$ is independent of $\tilde{x}$ and can be written as $H(c)$. In this section, we also extend the main result to $\tilde{x}$ having support on $\mathbb{R}$. The following lemma simplifies (\ref{xx1}) to obtain a straightforward characterization of the non-disclosure equilibrium price. 

\begin{lemma1}
\label{pd1}
The non-disclosure price $P^*(\emptyset)<\mu$ is given by the solution to $\Gamma(P^*(\emptyset))=0$, where:
\begin{equation}
\Gamma(y)\equiv \int x(1-H(x))  f(x+y)dx. \label{fin1}
\end{equation}
\end{lemma1}

\paragraph{}

The special case  $H(c)=1_{c\geq c_0}$ corresponds to a single mass point at $c_0$,
\begin{equation}\label{ver0}
\int_{-\infty}^{c_0} x f(x+P^*(\emptyset)) dx=0,
\end{equation}
offering a condensed characterization of the non-disclosure price in this type of model.\ To see why, denoting $\tau_0=c_0+P(\emptyset)$ as the corresponding disclosure threshold
and changing variables in (\ref{ver0}) yields
\begin{equation*}
0=\int_{-\infty}^{c_0} x f(x+\tau_0-c_0) dx=\int_{-\infty}^{\tau_0} (x-\tau_0+c_0) f(x) dx=\int_{-\infty}^{\tau_0} x f(x) dx-F(\tau_0)(\tau_0-c_0),
\end{equation*}
which, dividing both sides by $F(\tau_0)$, simplifies to the familiar indifference condition for the marginal discloser 
\begin{equation}\label{r0}
\mathbb{E}(x|x< \tau_0)=\tau_0-c_0.
\end{equation}

 A model in which the manager is informed with probability $q\in (0,1)$ can be similarly mapped to a function $H(c)=q \times \mathbf{1}_{c\geq 0}$, corresponding to a random cost that realizes at $c=0$ with probability $q$ and $c=\infty$ with probability $1-q$. Then
\begin{eqnarray*}
0&=&\int_{-\infty}^0 x f(x+P^*(\emptyset)) dx+ \left(1 - q\right) \int_0^{\infty} x f(x+P^*(\emptyset)) dx
\end{eqnarray*}
where, given that $P^*(\emptyset)=\tau^*$, the last equation simplifies to the equilibrium in \citeasnoun{junkwo88} in which $p = 1 - q$,
\begin{equation}\label{r1}
\left(1 - q \right)(\mu-\tau^*)= q\;\int_{-\infty}^0 F(x+\tau^*) dx.
\end{equation}
As these examples illustrate, while the equilibrium characterizations in  (\ref{r0}) and (\ref{r1}) may appear different, they are all special cases of the general condition $\Gamma(P^*(\emptyset))=0$ in (\ref{fin1}). An interesting property of (\ref{fin1}) is that, defining $\Delta\tilde{x} \equiv \tilde{x}-P^*(\emptyset)$ as the difference between fundamental and non-disclosure price, the equilibrium can be re-arranged as:
\begin{eqnarray*}
\mathbb{E}(\Delta \tilde{x}(1-H(\Delta \tilde{x})))=0, \quad \text{ equivalently, } \quad
\widehat{\mathbb{E}}(\Delta \tilde{x})=0,
\end{eqnarray*}
where the expectation $\widehat{\mathbb{E}}(.)$ is taken according to a modified probability measure $g( x)\equiv f( x)(1-H(\Delta x))/\int f(x')(1-H(\Delta x')) d x'$. In the special case of $H(c)=\mathbf{1}_{c=\infty}$, that is, the cost is infinite and no firm can disclose, this equation simplifies to $\mathbb{E}(\Delta \tilde{x})=0$ and the non-disclosure price is the unconditional mean because there is no discount for strategic disclosure.\ As the probability mass in $H(.)$ shifts toward lower disclosure costs, the modified probability  measure $g( x)$ assigns greater weight $(1-H(\Delta x))/\int (1-H(\Delta x')) d x'$ to correct for adverse selection and the non-disclosure price decreases. 

\begin{proposition1}\label{stat1}
If $H(0)=1$, there is no partial disclosure equilibrium.\ Otherwise, if $\tilde{x}$ has a thin lower tail, defined as $\lim_{x\rightarrow -\infty} e^{-a x}f(x)=0$ for any $a>0$, there exists a solution to (\ref{fin1}). If in addition $f(.)$ satisfies the monotone likelihood ratio property, this solution is unique.  \end{proposition1}

Thin lower tails are sufficient but not necessary for the existence of a partial disclosure equilibrium, which can be seen clearly in the case of a fixed cost. Remark~\ref{remark:fixed} provides a necessary and sufficient characterization for existence in this case and shows that, with thicker lower tails, a strictly positive disclosure cost may fail to generate any partial disclosure equilibrium.

\begin{remark1}\label{remark:fixed} 
Assume that the disclosure cost is fixed at $c_0>0$. Then $\tau$ is the threshold of a partial disclosure equilibrium if and only if

\begin{equation}\label{eq:fixed_cost}
\delta(\tau)\equiv \tau- \frac{\int_{-\infty}^\tau x d F(x)}{F(\tau)} =c_0.
\end{equation}
Consequently,
\begin{enumerate}
\item A partial disclosure equilibrium exists if and only if $c_0\ge \inf_{\tau: \,F(\tau)\in(0,1)}\delta(\tau)$.
\item If $\delta(\tau)$ is strictly increasing in $\tau$, the partial disclosure equilibrium is unique.
\item If $\inf_{\tau: \,F(\tau)\in(0,1)}\delta(\tau)>0$, there exists an interval of strictly positive costs $c_0\in\bigl(0,\inf_\tau \delta(\tau)\bigr)$ for which \eqref{eq:fixed_cost} admits no solution and no partial disclosure equilibrium exists.
\end{enumerate}
\end{remark1}

Item 3 in Remark \ref{remark:fixed} highlights that a positive cost may be insufficient to sustain partial disclosure. Intuitively, when the lower tail is thicker, the adverse selection discount embedded in $P(\emptyset)$ can become so severe that no partial disclosure equilibrium exists. Example \ref{ex:laplace} illustrates this mechanism in closed form for a Laplace fundamental which assigns more probability to extreme events.

\begin{example1}\label{ex:laplace}
Suppose $\tilde x$ is Laplace with density $f(x)=\frac{1}{2} e^{-|x|}$ on $\mathbb R$ and $\tilde c=c_0>0$ is fixed. Then
\begin{equation*}
\delta(\tau)= 
\begin{cases}
1 & \text{ if } \tau\le 0 \\
\frac{2\tau+e^{-\tau}}{2-e^{-\tau}} & \text{ if } \tau>0
\end{cases},
\end{equation*}
and $\delta(\tau)$ is strictly increasing on $(0,\infty)$ with $\lim_{\tau\to 0}\delta(\tau)=1$ and
$\lim_{\tau\to\infty}\delta(\tau)=\infty$. If $c_0<1$, there is no $\tau$ that solves \eqref{eq:fixed_cost} and no
partial disclosure equilibrium exists and if $c_0>1$, there exists a unique partial disclosure equilibrium such that the threshold $\tau^*>0$. Otherwise, that is, $c_0=1$, every $\tau\le 0$ supports a partial disclosure equilibrium and a continuum of equilibria exists.
\end{example1}

\textit{}

We discuss next additional comparative statics if the distribution of costs is normally distributed. In what follows, suppose  that the solution to (\ref{pd1}) is unique with $\Gamma'(P(\emptyset))<0$.\footnote{If the solution is not unique, it was shown in Section \ref{section3} that this derivative is non-negative in a partial disclosure neologism-proof equilibrium.}

\begin{proposition1}\label{rofo}
If $H(c)=\Phi((c-m)/\sigma)$, where $\Phi(.)$ is the c.d.f. of $N(0,1)$, then the probability of disclosure is decreasing in the average disclosure cost $m$. If $m$ is sufficiently small or $\sigma$ is sufficiently large, the probability of disclosure is decreasing in $\sigma$.
\end{proposition1}

The first claim in Proposition \ref{rofo} is intuitive, as it preserves the classic comparative statics that greater cost, in expectation, dissuade voluntary disclosure; this property is preserved even if costs are not deterministic. In addition, with additional structure imposed on the cost distribution (i.e., a normally distributed disclosure cost $\tilde{c}$), we demonstrate that cost uncertainty may have implications different from uncertainty about fundamentals; recall, as benchmark, that uncertainty $\tilde{x}$ about fundamentals does not affect the \textit{probability} of disclosure under uncertain information endowment \cite{junkwo88} and increases the probability of disclosure under costly disclosure \cite{ver90}. However, empirically, the evidence as to how uncertainty affects disclosure is mixed \cite{biljenlev15,hoplima23}, suggesting that the type of uncertainty may play an important role. Indeed, the comparative static is consistent with periods of considerable uncertainties about the consequences of disclosure - for example documented by \citeasnoun{hoplima23} during the COVID-19 outbreak - reducing incentives to disclose.

\section{Discrete Choice Estimation } \label{estimation}

\paragraph{}
The standard theories of \citeasnoun{ver83} and \citeasnoun{dye85} present an empirical challenge to be tested or quantified with real data. Specifically, these theories make the (counter-factual) prediction that firms never disclose below a certain threshold - a property rarely observed in practice. In fact, in most applications, the cross-sectional distribution of disclosures has full support and seems to feature frequent disclosures of ``bad" news \cite{ski94,kotshuwys09}.
Such empirical patterns are inconsistent with a strict sanitization equilibrium where firms omit sufficiently unfavorable information.
It may be tempting to ignore the obvious discrepancy between the model and data, that is, to view it as empirical unmodeled noise left outside of the assumed theory. However, such noise is not data error but about which news is actually being withheld - therefore, it affects the entire determination of the equilibrium as investors must rationally infer the correct non-disclosure price accounting for bad news disclosures. This rational process is exactly what the generalized framework with random costs is designed to capture.

\begin{figure}[h!]
\begin{center}
\includegraphics[scale=.7]{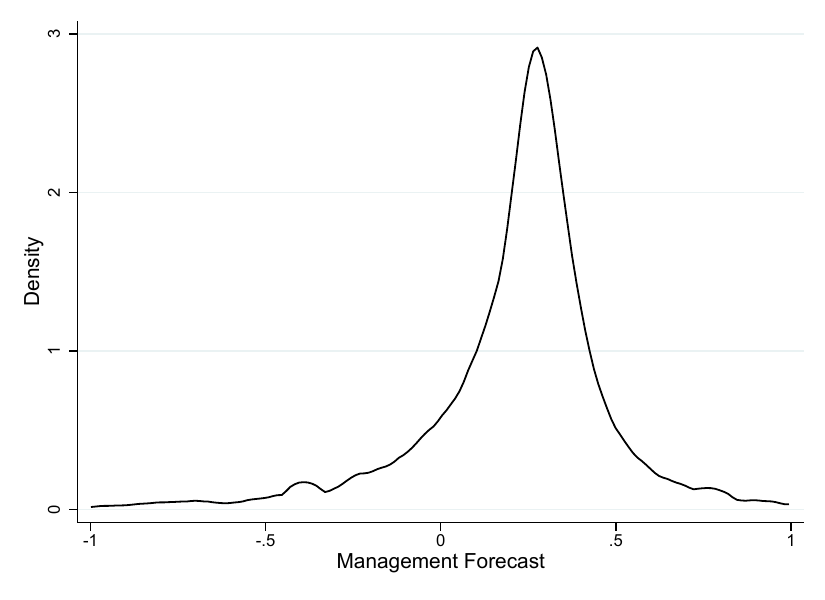}
\end{center}
\caption{Density plot of annual EPS management forecasts, from \protect\citeasnoun{bermamar20}} \label{nom1}
\end{figure}

To further illustrate this empirical fact, Figure \ref{nom1} presents the distribution of yearly earnings per share (EPS) forecasts  in \citeasnoun{bermamar20}. As in this earlier study, to create a homogeneous sample, all forecasts are standardized, subtracting the analyst consensus of the forecasted earnings before the earnings announcement, and dividing by the standard deviation of actuals. As a result, forecasts and realized earnings are defined as ``surprises" relative to the consensus and expressed in units of standard deviation of EPS. The average forecast is $0.21$, which suggests that firms withhold negative news but only \textit{in expectation}. Forecasts are widely distributed over all surprises, which cannot be rationalized without sufficient randomness in the friction.

Fortunately, the model can be estimated using standard reduced-form methods available in standard statistical packages without solving for the equilibrium. We denote the privately-observed information for an observation as $x_t$ and $d_t=x_t$ as a management forecast, with $c_t$ the current disclosure friction. Interestingly, this model presents a structure similar to discrete choice theory \cite{mcf72}. To begin with, let us denote an indicator variable $D_t=\mathbf{1}_{d_t\ne \emptyset}$ when the manager makes a forecast and assume that
\begin{eqnarray*}
\begin{pmatrix}
x_t\\c_t
\end{pmatrix}\sim \mathcal{N}\left(\begin{pmatrix}
0\\m_c
\end{pmatrix},\begin{pmatrix}
\sigma_x^2 & 0\\ 0 & \sigma_c^2
\end{pmatrix}\right),
\end{eqnarray*}
so that $m_c$ indicates the average cost, $\sigma_c$ the uncertainty about costs, and $\sigma_x$ the uncertainty about fundamentals.
Therefore,
\[
\Pr(D_t = 1 \mid x_t) = \mathbb{P}(c_t < x_t - P(\emptyset)\mid x_t) = \Phi\left(
\frac{x_t - P(\emptyset) - m_c}{\sigma_c}
\right),
\]
where \( \Phi(\cdot) \) denotes the cumulative distribution function of the standard normal distribution.

The above model could be estimated as a probit, using $D_t$ as a dependent variable, $x_t$ as an independent variable, and a latent disturbance term normalized to $\mathcal{N}(0,1)$.
However, this approach is infeasible because $x_t$ is not observed conditional on a non-disclosure. Hence, we consider a different approach using the fact that ex-post realizations (the actual forecasted EPS) are observed. Specifically, denote $y_{t+1}$ as the actual realized EPS (for the period forecasted by $x_t$). Under our maintained assumption that managers form rational priors, $x_t=\mathbb{E}_t(y_{t+1})$. A well-known property of conditional expectations is that
\begin{equation*}
y_{t+1}=x_t+u_t,
\end{equation*}
equal to the manager's expectation $x_t$ and an orthogonal innovation $u_t$ with variance $\sigma_u^2$. We are equipped to derive a feasible estimation procedure for the model using $\Pr(D_t=1|y_{t+1})$, that is, conditioning on the \textit{realized} EPS, which is observed even if the manager does not make a forecast.

To derive this probability, consider the joint distribution\[
\begin{pmatrix}
x_t \\
c_t \\
y_{t+1}
\end{pmatrix}
\sim \mathcal{N} \left(
\begin{pmatrix}
0 \\
m_c \\
0
\end{pmatrix},
\begin{pmatrix}
\sigma_x^2 & 0 & \sigma_x^2 \\
0 & \sigma_c^2 & 0 \\
\sigma_x^2 & 0 & \sigma_x^2 + \sigma_u^2
\end{pmatrix}
\right).
\]
Then the conditional distribution of \( (x_t, c_t) \mid y_{t+1} \) is given by
\begin{equation*}
\begin{pmatrix}
x_t \\
c_t
\end{pmatrix}
\Bigg| y_{t+1}
\sim
\mathcal{N} \left( 
\begin{pmatrix}
\frac{\sigma_x^2}{\sigma_x^2 + \sigma_u^2} y_{t+1} \\
m_c 
\end{pmatrix},
\begin{pmatrix}
\frac{\sigma_x^2 \sigma_u^2}{\sigma_x^2 + \sigma_u^2} & 0 \\
0 & \sigma_c^2 
\end{pmatrix}
\right).
\end{equation*}
Define $z_t \equiv x_t - c_t \). It follows that \( z_t \mid y_{t+1} \sim \mathcal{N}(\mu_z, \sigma_z^2) \), where
\begin{eqnarray*}
\mu_z \equiv \left( \frac{\sigma_x^2 }{\sigma_x^2 + \sigma_u^2} \right) y_{t+1} - m_c \quad \text{ and } \quad
\sigma_z^2 \equiv \frac{\sigma_x^2 \sigma_u^2}{\sigma_x^2 + \sigma_u^2}
+ \sigma_c^2.
\end{eqnarray*}
This implies the following probability of disclosure:
\begin{equation*}
\Pr(D_t = 1 \mid y_{t+1}) =  
\Phi\left(
\frac{
\left( \frac{\sigma_x^2 }{\sigma_x^2 + \sigma_u^2} \right) y_{t+1} - m_c - P(\emptyset)
}{\sigma_z}
\right).
\end{equation*}
The coefficients of a probit $\Pr(D_t = 1 \mid y_{t+1}) =  
\Phi(\alpha_0+\alpha_1 y_{t+1})$ are \begin{eqnarray}\label{A1}
\alpha_0 &=& \frac{- (m_c + P(\emptyset))}{\sigma_z}\\
\alpha_1 &=& \frac{\sigma_x^2 }{(\sigma_x^2 + \sigma_u^2)\sigma_z}.\label{A2}
\end{eqnarray}

Table \ref{probit} reports the results of the probit regression. Consistent with the model, the probability of disclosure increases with higher realized earnings. Further, the intercept is negative, indicating that the manager is more likely not to disclose absent an earnings surprise. Note that this property is also inconsistent with pure information endowment models, as in \citeasnoun{dye85} and \citeasnoun{junkwo88}, since the manager tends to frequently withholds news with small positive earnings surprises.

\begin{table}[htbp]
\centering
\renewcommand{\arraystretch}{1.3}
\begin{tabular}{lccc}
\toprule
\textbf{Variable} 
& \textbf{Coefficient} 
& \textbf{Std. Error} 
& \textbf{95\% CI} \\
\midrule
$\text{EPS}$     & 0.1032 & 0.0096 & [0.0843,\; 0.1221] \\
Constant             & -0.4084 & 0.0094 & [-0.4268,\; -0.3900] \\
\bottomrule
\end{tabular}
\vspace{0.75em}
\begin{minipage}{0.9\textwidth}
\footnotesize
\textit{Notes:} This table reports maximum-likelihood estimates from the probit model $Pr(D_t=1\mid EPS_t)=\Phi(\beta_0+\beta_1\,EPS_t)$,
where $D_t=\mathbf{1}\{d_t\neq \emptyset\}$ is an indicator equal to one if the firm issues a management earnings forecast at the earnings announcement for next year's earnings (and zero otherwise), and $\Phi(\cdot)$ is the standard normal c.d.f. The regressor $EPS_t$ denotes the realized earnings-per-share  for the forecasted period expressed as a standardized earnings surprise (actual EPS minus the pre-announcement analyst consensus forecast, scaled by the sample standard deviation of actual EPS).  The 95\% confidence interval is computed as estimate $\pm 1.96\times$ standard error. Estimates are based on $N=19{,}060$ firm-year observations. \end{minipage}
\caption{Probit Regression Results \label{probit}}
\end{table}

To interpret the results of the probit, it is necessary to infer from the estimated reduced-form coefficients the five unknown structural parameters  $(P(\emptyset),\sigma_x,\sigma_u,m_c,\sigma_c)$, which requires to use a second stage. The coefficient $\sigma_u$ can be directly estimated as the standard deviation $\hat{\sigma}_u$ of $y_{t+1}-x_t$ over all disclosing firms, given that $u_t$ is independent of $x_t$ and, then, $\sigma_x$ can be readily estimated as $\hat{\sigma}_x=\sqrt{\hat{\sigma}_y^2-\hat{\sigma}_u^2}$, where $\hat{\sigma}_y$ is the standard errors of realized EPS. Next, similar to \citeasnoun{cheliu20}, the non-disclosure price $P(\emptyset)$ must satisfy the law of total expectations,
\begin{equation*}
\mathbb{E}(D_t)\mathbb{E}(x_t|D_t=1)+\mathbb{E}(1-D_t)P(\emptyset)=\mathbb{E}(x_t)=0,
\end{equation*}
which implies that the non-disclosure price can be estimated as
\begin{equation*}
\hat{P}(\emptyset)=-\frac{\hat{p}_d}{1-\hat{p}_d}\hat{m}_d,
\end{equation*}
where $\hat{p}_d$ indicates the disclosure frequency and $\hat{m}_d$ is the average disclosure. Incorporating these moments into (\ref{A1})-(\ref{A2}) and solving for the remaining structural parameters yields the following second-stage to obtain the characteristics of the random cost:
\begin{eqnarray*}
m_c = \frac{p_d}{1 - p_d} m_d -\frac{\alpha_0}{\alpha_1}\frac {\sigma_y^2 - \sigma_u^2} {\sigma_y^2} \quad \text{ and } \quad
\sigma_c^2 = \frac{(\sigma_y^2 - \sigma_u^2) \bigl( (1 - \alpha_1^2 \sigma_u^2)\sigma_y^2-\sigma_u^2 \bigr)} {\alpha_1^2 \sigma_y^4}.
\end{eqnarray*}
We run this analysis on the management forecast data used in Figure \ref{nom1}. 
\begin{table}[htbp]
\centering
\label{tab:bootstrap_results}
\renewcommand{\arraystretch}{1.3}
\begin{tabular}{llcccc}
\toprule
\textbf{Param.} 
& \textbf{Definition}
& \textbf{Estimate} 
& \textbf{Std. Err.} 
& \textbf{CI Lower} 
& \textbf{CI Upper} \\
\midrule
$\sigma_u$
& Std.\ dev.\ of earnings innovation $u_t$
& 0.7620  & 0.0114  & 0.7397  & 0.7843 \\

$\sigma_x$
& Std.\ dev.\ of manager expectation $x_t$
& 0.6476  & 0.0143  & 0.6196  & 0.6756 \\

$P(\emptyset)$
& Non-disclosure price
& -0.1111 & 0.0026  & -0.1162 & -0.1060 \\

$ m_c$
& Mean disclosure cost
& 1.7714  & 0.1497  & 1.4780  & 2.0649 \\

$\sigma_c$
& Std.\ dev.\ of disclosure cost
& 4.0350  & 0.3603  & 3.3288  & 4.7412 \\
\bottomrule
\end{tabular}

\vspace{0.75em}
\begin{minipage}{0.9\textwidth}
\footnotesize
\textit{Notes:} This table reports structural estimates implied by the random-cost disclosure model. All earnings variables are measured as standardized EPS surprises (actual EPS minus the pre-announcement analyst consensus forecast, scaled by the sample standard deviation of actual EPS), so $P(\emptyset)$, $m_c$, and the reported standard deviations are expressed in the same standardized units.
The reported estimates come from a two-stage procedure: (i) a probit for $D_t$ on realized earnings (Table~\ref{probit}); (ii) moment restrictions and the probit mapping to recover $(P(\emptyset),\sigma_x,\sigma_u,m_c,\sigma_c)$. ``Std. Err.'' is the bootstrap standard error from re-estimating both stages. ``CI Lower'' and ``CI Upper'' report 95\% confidence intervals computed as estimate $\pm 1.96\times$ (bootstrap standard error). Bootstrap uses 1{,}000 replications; sample size is $N=19{,}060$.
\end{minipage}

\caption{Structural Estimates}
\end{table}

This identification strategy does not carry over to models in which fundamentals and costs are correlated, in which case the model is only set-identified and identification would typically require an exogenous variation. To see why, consider the joint distribution\[
\begin{pmatrix}
x_t \\
c_t \\
y_{t+1}
\end{pmatrix}
\sim \mathcal{N} \left(
\begin{pmatrix}
0 \\
m_c \\
0
\end{pmatrix},
\begin{pmatrix}
\sigma_x^2 & \rho \sigma_x \sigma_c & \sigma_x^2 \\
\rho \sigma_x \sigma_c & \sigma_c^2 & \rho \sigma_x \sigma_c \\
\sigma_x^2 & \rho \sigma_x \sigma_c & \sigma_x^2 + \sigma_u^2
\end{pmatrix}
\right),
\]
where $\rho$ now indicates the correlation between fundamentals and costs. Then the conditional distribution of \( (x_t, c_t) \mid y_{t+1} \) is given by
\begin{equation*}
\begin{pmatrix}
x_t \\
c_t
\end{pmatrix}
\Bigg| y_{t+1}
\sim
\mathcal{N} \left( 
\begin{pmatrix}
\frac{\sigma_x^2}{\sigma_x^2 + \sigma_u^2} y_{t+1} \\
m_c + \frac{\rho \sigma_x \sigma_c}{\sigma_x^2 + \sigma_u^2} y_{t+1}
\end{pmatrix},
\begin{pmatrix}
\frac{\sigma_x^2 \sigma_u^2}{\sigma_x^2 + \sigma_u^2} & \text{Cov}(x_t, c_t \mid y_{t+1}) \\
\text{Cov}(x_t, c_t \mid y_{t+1}) & \sigma_c^2 - \frac{(\rho \sigma_x \sigma_c)^2}{\sigma_x^2 + \sigma_u^2}
\end{pmatrix}
\right),
\end{equation*}
where
\begin{eqnarray*}
\text{Cov}(x_t, c_t \mid y_{t+1}) &=& \rho \sigma_x \sigma_c - \frac{\rho \sigma_x^3 \sigma_c}{\sigma_x^2 + \sigma_u^2}.
\end{eqnarray*}
Define $z_t \equiv x_t - c_t \). It follows that \( z_t \mid y_{t+1} \sim \mathcal{N}(\mu_z, \sigma_z^2) \), where
\begin{eqnarray*}
\mu_z &=& \mathbb{E}(z_t \mid y_{t+1}) = \left( \frac{\sigma_x^2 - \rho \sigma_x \sigma_c}{\sigma_x^2 + \sigma_u^2} \right) y_{t+1} - m_c\\
\sigma_z^2 &=& \text{Var}(x_t \mid y_{t+1}) + \text{Var}(c_t \mid y_{t+1}) - 2\,\text{Cov}(x_t, c_t \mid y_{t+1})\nonumber\\
&=& \frac{\sigma_x^2 \sigma_u^2}{\sigma_x^2 + \sigma_u^2}
+ \sigma_c^2 \left( 1 - \frac{\rho^2 \sigma_x^2}{\sigma_x^2 + \sigma_u^2} \right)
- 2\rho \sigma_x \sigma_c  \left( 1- \frac{\sigma_x^2}{\sigma_x^2 + \sigma_u^2} \right).
\end{eqnarray*}
This implies the following probability of disclosure:
\begin{equation*}
\Pr(D_t = 1 \mid y_{t+1}) =  
\Phi\left(
\frac{
\left( \frac{\sigma_x^2 - \rho \sigma_x \sigma_c}{\sigma_x^2 + \sigma_u^2} \right) y_{t+1} - m_c - P(\emptyset)
}{\sigma_z}
\right).
\end{equation*}
The coefficients of a probit $\Pr(D_t = 1 \mid y_{t+1}) =  
\Phi(\alpha_0+\alpha_1 y_{t+1})$  are
\begin{eqnarray*}
\alpha_0 = \frac{- (m_c + P(\emptyset))}{\sigma_z} \quad \text{ and } \quad
\alpha_1 = \frac{\sigma_x^2 - \rho \sigma_x \sigma_c}{(\sigma_x^2 + \sigma_u^2)\sigma_z}.
\end{eqnarray*}

\begin{figure}[h!]
\raisebox{-0.37\height}{\includegraphics[scale=.6]{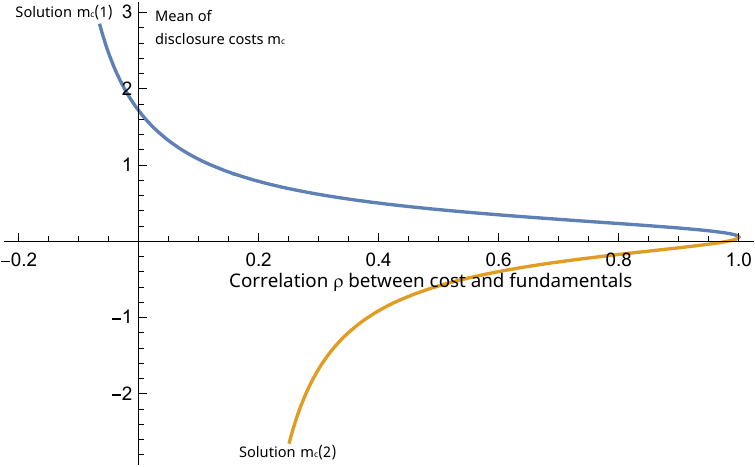}} \hspace{.1cm}
\includegraphics[scale=.6]{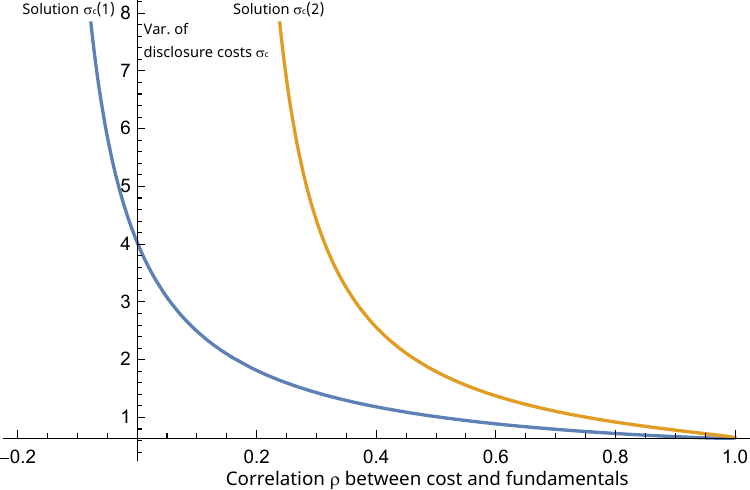}
\caption{Estimated Costs, with varying correlation between cost and fundamentals}\label{GR1}
\end{figure}

In the above equation, the correlation coefficient $\rho$ constitutes an additional free parameter in a (previously) exactly identified model where the number of parameters was equal to the number of equations. As a result, the structural parameters are identified only conditional on a given value of $\rho\in(-1,1)$. Figure~\ref{GR1} illustrates this relationship by plotting the implied mean and variance of disclosure costs as functions of $\rho$, and derives the implies values of $m_c$ (left) and $\sigma_c^2$ (right). For sufficiently negative correlations (approximately $\rho<-0.15$), the model admits no solution for $\sigma_c$, allowing us to rule out such values of $\rho$ and implying that the model is set-identified. For intermediate values of $\rho$, there exists a unique solution for $(m_c,\sigma_c^2)$, which in turn uniquely pins down the remaining parameters $(\sigma_u^2,\sigma_x^2,P(\emptyset))$. However, these estimates vary with $\rho$, which itself is not identified from the data. For larger positive correlations, the model may admit multiple solutions for $(m_c,\sigma_c^2)$ even for a given $\rho$. The source of this multiplicity is that strong positive correlation between fundamentals and disclosure costs can generate similar upper-tail behavior in observed forecasts through distinct mechanisms: either a highly dispersed distribution of forecasts combined with, on average, negative disclosure costs, or a less dispersed forecast distribution paired with higher average costs. Both configurations imply comparable probabilities of observing extreme realizations. 

\section{Conclusion}\label{conclusion}

\paragraph{}
This paper offers a unified treatment of many classic models of disclosure, integrating all frictions into a single object: a specification of privately observed disclosure costs, realized only when the firm chooses to disclose. The framework encompasses most commonly-used models of voluntary disclosure and yields generalizable intuitions that can be tested empirically, without requiring the selection of specific, and potentially unobservable, aspects of the theory. In particular, we prove that disclosure is decreasing in the frictions, increasing in fundamental uncertainty, and that greater uncertainty about disclosure costs reduces disclosure when the cost uncertainty is sufficiently large. Put differently, the framework provides a novel perspective on evidence games, akin to noisy signaling models, where the firm observes two private signals, fundamentals and costs, but can only communicate a single-dimensional information (see, e.g., \citeasnoun{fisver00} and \citeasnoun{dyesri04} in the costly lying literature).

Two critical simplifying assumptions need to be emphasized, suggesting that the current framework may represent a special case of an even broader theory. First, we restrict attention to binary disclosure decisions: firms either reveal their private information or remain silent. A more general message space introduces substantial challenges related to off-equilibrium beliefs;  these games with richer message spaces often cannot be resolved through simple threshold strategies. Models such as \citeasnoun{kir97}, \citeasnoun{pae05}, \citeasnoun{dyefin07}, and \citeasnoun{aghsmi23} provide important insights into these complexities, although fully incorporating all considerations discussed here remains highly non-trivial.

Second, the models considered here focus on (myopic) price maximization, where the manager values only the posterior beliefs induced by disclosure. In many settings, however, managers may retain a stake in the asset being sold, introducing an alternative channel for signaling through various ownership or investment decisions tied to their private information. In such cases, an equilibrium may be partially revealing if managers credibly signal their private information by, for example, retaining a larger share or investing more when their private information is favorable \cite{kansinspe05,beygut12}. We do not explicitly address competition between different communication channels in this paper.

Overall, by reconciling separate streams of the literature, the framework provides new theoretical and empirical tools for studying strategic voluntary disclosures. In doing so, we highlight different paths that could likely extend our current knowledge over the communication of evidence.

\section*{Appendix}

\paragraph{}

\textbf{Proof of Proposition \ref{unr}:} For the ``if" part, suppose that, with probability one, $\tilde{x}-\tilde{c}\geq 0$. Then a non-disclosure is off-equilibrium and the belief $P(\emptyset)=0$ is consistent with Definition \ref{J1}. For the ``only if" part, unraveling would imply that disclosure occurs when $\tilde{x}-\tilde{c}\geq P(\emptyset)= 0$. \qed

\textbf{Proof of Proposition \ref{prop:money_burning}:}
We construct an equilibrium as follows by specifying (i) the manager's strategy and (ii) investors' beliefs and pricing: (i) After observing $(x,c)$, the manager chooses $y(x,c)=\min\{x,c\}$. If $c\le x$, then $y=c\ge c$ and verification succeeds, so investors observe $x$. If $c>x$, then $y=x<c$ and verification fails, so investors observe only $y$. (ii) Upon successful verification, investors observe $x$ and set the competitive price $P=x$. Upon failed verification with observed expenditure $y$, investors believe that $x=y$ with probability one and set
the price $P = y$.

We first show that the manager has no profitable deviation. Fix a type $(x,c)$. Suppose that $c\le x$. Under the proposed strategy the manager chooses $y=c$. Then verification succeeds and the manager obtains a payoff $P-y=x-c\ge 0$. Consider a deviation to some $y'$. If $y'<c$, verification fails and the manager's payoff is $y'-y'=0\le x-c$. If $y'\ge c$, verification succeeds and the manager's payoff is $x-y'\le x-c$. Hence no deviation yields a strictly higher payoff. Suppose that $c>x$. Under the proposed strategy the manager chooses $y=x$. Then verification fails and the manager obtains a payoff $P - y = x - x = 0$. If the manager deviates to some $y'<c$, verification still fails and the payoff remains zero. If the manager deviates to some $y'\ge c$, verification succeeds but the payoff is $x-y'<0$, because $y'\ge c>x$. Hence, there is no profitable deviation.

Next, we verify the consistency condition. Given the strategy, when investors observe failed verification and expenditure $y$, the only type who sends this message is $y=x$ which is less than some $c$. Hence, the belief $x=y$ is consistent with Bayes rule and competitive pricing yields $P(y)=\mathbb{E}[x| y]=y$ after failed verification, and $P=x$ after disclosure. 

So the proposed strategies, beliefs, and prices constitute a perfect Bayesian equilibrium. In the equilibrium, the market price equals $x$ both after disclosure, because $x$ is observed, and after failed verification, because $P(y)=y=x$. Therefore, the equilibrium is fully revealing. \qed

\textbf{Proof of\ Proposition \ref{exist}:} Since existence is already shown under the condition of Proposition \ref{unr}, suppose that neither type of equilibrium exists. Then, denoting $\overline{H}(c|x)\equiv 1-H(c|x)$ and the continuous function
$$\psi(y)=\frac{\int x\overline{H}(x - y|x)dF(x) }{\int \overline{H}(x - y|x)dF(x)}-y.$$
If there is no unraveling equilibrium, non-disclosure occurs with positive probability at $y=0$, hence $\psi(0)>0$. Further, evaluate the right-hand side at $y=\infty$. It follows from $\overline{H}(x-y|x) \to 1$ as $y \to \infty$ that
$$\lim_{y\rightarrow \infty}\psi(y)=\lim_{y\rightarrow \infty}\frac{\int xdF(x) }{\int dF(x)}-y=-\infty.$$
Therefore, there exists at least one solution $y$ to $\psi(y)=0$ by the intermediate value theorem.

We prove next the second part of the Proposition. By~\eqref{xx}, the non-disclosure price $y>0$ in any partial-disclosure equilibrium is a root of
\begin{equation*}
G(y)\equiv \int (x-y)\,\overline{H}(x-y| x)\,dF(x)=0.
\end{equation*}
Under $H(x| x)=1$ we have $\overline{H}(x| x)=0$ and thus $G(0)=0$.
Moreover, since $\overline{H}(x-y| x)\to 1$ as $y\to\infty$, we have $\lim_{y\to\infty}G(y)=\mu-y<0$ and hence $G(y) < 0$ for $y$ sufficiently large by continuity.

It is therefore sufficient to show that $G(y)>0$ for some $y>0$ small. Note that
\begin{equation*}
G'(y)=\int\left(-\overline{H}(x-y| x)+(x-y)\,h(x-y| x)\right)\,dF(x),
\end{equation*}
so that
\begin{equation*}
G'(0)=\int x\,h(x\mid x)\,dF(x)=0
\end{equation*}
under the condition that $h(x| x)=0$. The second derivative can be computed similarly as
\begin{equation*}
G''(0)= -\int x\,h_1(x| x)\,dF(x)>0,
\end{equation*}
where $h_1(c| x)\equiv \partial h(c| x)/\partial c$ and the inequality follows from
$h_1(x\mid x)<0$.
A Taylor expansion around $0$ then implies $G(y)>0$ for $y>0$ small. By continuity and since $G(y)<0$ for $y$ sufficiently large, there exists at least one root $y^*>0$, which corresponds to a partial disclosure equilibrium.
\qed

\textbf{Proof of Proposition \ref{prop:unique}:} By Proposition \ref{unr}, there exists an unraveling equilibrium if and only if $\tilde{x}-\tilde{c}\geq 0$ almost surely. We rule out existence of a partial disclosure equilibrium in case (i) where this condition holds, while show that there is a unique partial disclosure equilibrium in case (ii) where this condition fails (recall that existence has been established in Proposition \ref{exist}).

Consider case (i) where $\tilde{x} - \tilde{c} \ge 0$ almost surely. Note that $\Lambda(y)$ is the right-hand side of equations~\eqref{xx1} and~\eqref{xx} where $y = P(\emptyset)$. There is a partial disclosure equilibrium if and only if there exists $y > 0$ such that $\Pr(\tilde{x} - \tilde{c} < y) > 0$ and $y = \Lambda(y)$, where $y$ is the non-disclosure price. Clearly, if $\Lambda(y) < y$ for all $y > 0$ such that $\Pr(\tilde{x} - \tilde{c} < y) > 0$, unraveling is the unique equilibrium. We next show necessity. Assume that unraveling is the unique equilibrium. As defined in the proof of Proposition \ref{exist}, $\psi(y) \equiv \Lambda(y) - y$. By hypothesis, there is no partial disclosure equilibrium. So $\psi(y)\neq 0$
for all $y>0$ such that $\Pr(\tilde x-\tilde c<y)>0$. Because $\psi(y) < 0$ for $y$ sufficiently large, if there exists $\hat{y} > 0$ with $\Pr(\tilde{x}-\tilde{c}< \hat{y}) > 0$ and $\psi(\hat{y})>0$, there must exist $y_0 > \hat{y}$ such that $\psi(y_0) = 0$ by continuity and the intermediate value theorem. The contradiction that $y_0$ is the non-disclosure price of a partial disclosure equilibrium implies that $\psi(y) < 0$ (i.e., $\Lambda(y) < y$) for all $y > 0$ such that $\Pr(\tilde{x} - \tilde{c} < y) > 0$.

Consider case (ii) where $\Pr(\tilde{x} - \tilde{c} \ge 0) < 1$. Then there is no unraveling equilibrium by Proposition \ref{unr}. It follows from Proposition \ref{exist} that there exists at least one partial disclosure equilibrium. Then the equilibrium is unique if $\psi(y)$ is monotonically decreasing in $y \in [0, \mu]$. Note that
\begin{eqnarray*}
    \psi'(y) & = & \frac{\int x h(x-y|x) d F(x) \int \overline{H}(x-y|x) d F(x) - \int h(x-y|x) d F(x) \int x \overline{H}(x-y|x) d F(x)}{\left(\int \overline{H}(x-y|x) d F(x) \right)^2} - 1 \\ & = & \frac{\int\left(x - \Lambda(y) \right) h(x-y|x) d F(x) }{\int \overline{H}(x-y|x) d F(x)} - 1. 
\end{eqnarray*}
If $\sup_{y \in [0, \mu]} \frac{\int \left(x - \Lambda(y)\right) h(x - y|x) d F(x)}{\int \overline{H}(x - y|x) d F(x)} < 1$, then $\psi'(y) < 0$ for all $y \in [0, \mu]$, which proves uniqueness of a partial disclosure equilibrium. 

Moreover, suppose that $\tilde{x}$ and $\tilde{c}$ are independent, i.e., $H(.| x)=H(.)$ for all $x$. We first consider the case of a fixed cost $\tilde{c} \equiv c_0$. Let $Q(y) \equiv \mathbb{E}(\tilde{x}|\tilde{x} \le y)$. Then $\Lambda(y) = Q(y + c_0)$ and $\Lambda'(y) = Q'(y + c_0)$. Further note that
\begin{equation*}
    Q'(y) = \frac{\int_{0}^{y} x f(x)\,dx}{F(y)} = \frac{y f(y) F(y) - \left( \int_{0}^{y} x f(x)\,dx\right) f(y)}{F(y)^2} = \frac{f(y)}{F(y)} \left( y - Q(y)\right).
\end{equation*}
It follows that
\begin{equation*}
\psi'(y) = \Lambda'(y) - 1 = Q'(y + c_0) - 1 = \frac{f(y + c_0)}{F(y+ c_0)} \left( y + c_0 - Q(y + c_0)\right) - 1 = \frac{f(y + c_0)}{F(y+ c_0)} \left( y + c_0 - \Lambda(y)\right) - 1.
\end{equation*}
Because $y = \Lambda(y)$ where $y = P(\emptyset)$ in equilibrium, we have $\psi'(P(\emptyset)) = c_0 \frac{f(P(\emptyset) + c_0)}{F(P(\emptyset) + c_0)} - 1$. If there is more than one $y$ such that $y = \Lambda(y)$, at least one of them must satisfy $\psi'(y) \ge 0$. But for all $y \ge 0$, $c_0 \frac{f(y + c_0)}{F(y + c_0)} < 1$ by assumption. So $\psi'(y) < 0$ for all $y$ such that $y = \Lambda(y)$. Hence, there is at most one equilibrium, which proves uniqueness.

Finally, we show that the partial disclosure equilibrium is unique if the density $f$ is strictly positive and logconcave on $\mathbb{R}_+$, still assuming that costs are type-independent. The equilibrium condition $y = \mathbb{E}\left(\tilde{x}|\tilde{x} - \tilde{c} < y \right)$ is equivalent to
\begin{equation} \label{propeq:unique}
    0 = \mathbb{E}\left(\tilde{x} - y|\tilde{x} - \tilde{c} < y \right).
\end{equation}
Let $\tilde{z} = \tilde{x} - y$. Conditional on non-disclosure, the density of $\tilde{z}$ is $\frac{f(x) \overline{H}(x - y)}{\int f(x') \overline{H}(x' - y) d x'} = \frac{f(z + y) \overline{H}(z)}{\int f(z' + y) \overline{H}(z') d z'}$. We claim that $\mathbb{E}_y \left(\tilde{z} \right) \equiv \mathbb{E} \left(\tilde{x} - y|\tilde{x} - \tilde{c} < y \right)$ is decreasing in $y$, where $\mathbb{E}_y\left(. \right)$ denotes the expectation under the density of $\tilde{z}$ when the non-disclosure price is $y$. Consider two non-disclosure prices $y_2 > y_1$. The likelihood ratio satisfies
\begin{equation*}
    \frac{f(z + y_2) \overline{H}(z) \int f(z' + y_1) \overline{H}(z') d z'}{f(z + y_1) \overline{H}(z) \int f(z' + y_2) \overline{H}(z') d z'} = \frac{f(z + y_2)}{f(z + y_1)} \frac{\int f(z' + y_1) \overline{H}(z') d z'}{\int f(z' + y_2) \overline{H}(z') d z'}. 
\end{equation*}
Let $l(z) \equiv \log f(z + y_2) - \log f(z + y_1)$. Because $f$ is logconcave, $\left( \log f \right)'$ is non-increasing and
\begin{equation*}
l'(z) = \left(\log f\right)'(z + y_2) - \left(\log f\right)'(z + y_1) \le 0.
\end{equation*}
So $\frac{f(z + y_2)}{f(z + y_1)}$ is non-increasing in $z$. It follows that the ratio of densities is non-increasing in $z$ (i.e., the monotone likelihood ratio property holds). Because the monotone likelihood ratio property implies that $\tilde{z}$ with non-disclosure price $y_2$ is first order stochastically dominated by $\tilde{z}$ with non-disclosure price $y_1$, it must be that $\mathbb{E}_{y_2} \left(\tilde{z} \right) \le \mathbb{E}_{y_1} \left(\tilde{z} \right)$, where the inequality is strict if the distributions are not identical. This implies that $\mathbb{E}_y \left(\tilde{z} \right)$ is decreasing in $y$. So Equation~\eqref{propeq:unique} $\mathbb{E}_y \left( \tilde{z} \right) = 0$ has at most one solution, which completes the proof that the partial disclosure equilibrium is unique.
\qed

\textbf{Proof of Proposition \ref{exist0}:}
Slightly abusing the notation, let $\mathcal{N}$ be the set of non-disclosure prices in the partial disclosure equilibria: $\mathcal N \equiv \{P\in\mathbb R_+:\;
P=\mathbb{E}(\tilde x| \tilde x-\tilde c < P)\}$. Take any equilibrium price $P_1\in\mathcal{N}$ such that $P_1 \notin \max_{\mathcal{N}} P$. Then there exists $P_2\in\mathcal N$ such that $P_2>P_1$. Augment the message space with a neologism message $m$ that is not used in equilibrium, and suppose that sending $m$ incurs an arbitrarily small private cost $\varepsilon>0$. Consider the candidate interpretation $P(m)=P_2$.

Under the equilibrium with non-disclosure price $P_1$, a type $(x,c)$ obtains payoff $\max\{P_1,x-c\}$. If investors interpret $m$ as $P_2$, then the payoff from deviating to $m$ is $P_2-\varepsilon$. It follows that (a) If $x-c\ge P_2$, then $\max\{P_1,x-c\}=x-c\ge P_2>P_2-\varepsilon$, so $(x,c)$ does not send $m$; (b) If $x-c<P_2$, then $\max\{P_1,x-c\}\le P_2$ and, for all such types except possibly a null set at
equality, $P_2-\varepsilon>\max\{P_1,x-c\}$, so these types strictly prefer sending $m$. Therefore, under the interpretation $P(m)=P_2$, the set of types who send $m$ is exactly $\{(x,c): x-c<P_2\}$. Hence, investors who observe $m$ rationally infer that the sender's type satisfies
$\tilde x-\tilde c<P_2$, and therefore set $P(m)=\mathbb E(\tilde x| \tilde x-\tilde c<P_2)$. Because $P_2\in\mathcal N$, it satisfies $P_2=\mathbb E(\tilde x| \tilde x-\tilde c<P_2)$. So the proposed interpretation $P(m)=P_2$ is consistent with rational expectations.

By Definition~\ref{def:np}, the existence of such a credible neologism $m$ implies that the equilibrium with non-disclosure price $P_1$ is not neologism-proof. Therefore, any neologism-proof equilibrium must have a non-disclosure price that is maximal in $\mathcal{N}$, i.e., features the lowest probability of disclosure.
\qed

\textbf{Proof of Corollary \ref{exist012}:} If a partial-disclosure equilibrium exists, there exists at least one Nash equilibrium with
a strictly positive non-disclosure price $P_2(\emptyset)>0$. An unraveling equilibrium corresponds to the lowest possible non-disclosure price, because non-disclosure is off equilibrium path and supported by $P_1(\emptyset)=0$. By Proposition \ref{exist0}, any equilibrium with non-disclosure price $P_1(\emptyset)$ that is not maximal among Nash equilibria is not neologism-proof. Because $P_2(\emptyset)>P_1(\emptyset)$, the unraveling equilibrium fails neologism-proofness whenever a partial disclosure equilibrium exists.
\qed

\textbf{Proof of Corollary \ref{cor:cs}:}
When $\Pr(\tilde x-\tilde c^\alpha<y)>0$, the equilibrium non-disclosure price satisfies
\begin{equation*}
P(\emptyset) = \mathbb{E}(\tilde x| \tilde x-\tilde c^\alpha<P(\emptyset))
=\frac{\int x \left(1 - H_\alpha(x-P(\emptyset)| x) \right)\,dF(x)}{\int \left(1 - H_\alpha(x-P(\emptyset)| x) \right)\,dF(x)}.
\end{equation*}
Let $\overline{P}_\alpha(\emptyset)$ be the maximal non-disclosure price in equilibrium.

\noindent Part (i): Let $r(x, y)\ \equiv\ \frac{1 - H_{\alpha'}(x-y| x)}{1 - H_\alpha(x-y| x)}$. This function is nondecreasing in $x$ as defined. Let $\nu_\alpha^y$ be the probability measure with density proportional to $\left(1 - H_\alpha(x-y| x) \right)\,dF(x)$: $d\nu_\alpha^y(x) = \frac{1 - H_\alpha(x-y| x)}{\int \left(1- H_\alpha(x'-y| x)\right)\,dF(x')}\,dF(x)$. Then $\mathbb{E}(\tilde{x}| \tilde{x}-\tilde{c}^\alpha<y)=\mathbb{E}_{\nu_\alpha^y}(x)$ and
\begin{equation*}
\mathbb{E}(\tilde{x}| \tilde{x}-\tilde{c}^{\alpha'}<y)=\frac{\mathbb{E}_{\nu_\alpha^y}(x\,r(x,y))}{\mathbb{E}_{\nu_\alpha^y}(r(x,y))}
=\mathbb{E}_{\nu_\alpha^y}(x)+\frac{Cov_{\nu_\alpha^y}(x,r(x,y))}{\mathbb{E}_{\nu_\alpha^y}(r(x,y))}.
\end{equation*}
Because $x$ and $r(x,y)$ are co-monotone under $\nu_\alpha^y$, Chebyshev's inequality implies
$Cov_{\nu_\alpha^y}(x,r(x,y))\ge 0$. It follows that
\begin{equation}\label{eq:TH_ge_TL}
\mathbb{E}(\tilde{x}| \tilde{x}-\tilde c^{\alpha'}<y)\ \ge\ \mathbb{E}(\tilde{x}| \tilde{x}-\tilde c^\alpha<y).
\end{equation}
Recall that $\psi_\alpha(y)\equiv \mathbb{E}(\tilde{x}| \tilde{x}-\tilde c^\alpha<y) - y$. Because $\psi_{\alpha'}(y)\ge \psi_\alpha(y)$ for all $y$ by \eqref{eq:TH_ge_TL}, $\psi_{\alpha'}(\overline{P}_\alpha(\emptyset))\ge 0$.

Moreover, as $y\to\infty$, we have $x-y\to -\infty$ and thus $1- H_\alpha(x-y| x)\to 1$. It follows that $\mathbb{E}(\tilde x| \tilde x-\tilde c^\alpha<y)\to \mu$ and $\psi_\alpha(y)=\mathbb{E}(\tilde x| \tilde x-\tilde c^\alpha<y) - y \to -\infty$. By continuity, there exists at least one solution to $\psi_{\alpha'} = 0$ at some $y \ge \overline{P}_\alpha(\emptyset)$. Hence $\overline{P}_{\alpha'}(\emptyset)\ \ge\ \overline{P}_\alpha(\emptyset)$. 

Finally, the disclosure probability in the environment $\alpha$ is
\begin{equation*}
D_\alpha
=\Pr(\tilde x-\tilde c^\alpha\ge \overline{P}_\alpha(\emptyset))
=\int H_\alpha(x-\overline{P}_\alpha(\emptyset)| x)\,dF(x).
\end{equation*}
By $\overline{P}_{\alpha'}(\emptyset)\ \ge\ \overline{P}_\alpha(\emptyset)$, it must be that
\begin{equation} \label{cor_eq1}
H_{\alpha'}(x-\overline{P}_{\alpha'}(\emptyset)| x)\ \le\ H_{\alpha'}(x-\overline{P}_\alpha(\emptyset)| x).
\end{equation}
Further by first-order stochastic dominance, we have
\begin{equation} \label{cor_eq2}
H_{\alpha'}(x-\overline{P}_\alpha(\emptyset)| x)\ \le\ H_\alpha(x-\overline{P}_\alpha(\emptyset)| x).
\end{equation}
Combining~\eqref{cor_eq1} and~\eqref{cor_eq2}, and integrating over $F$, we show that $D_{\alpha'} \le D_{\alpha}$.

\noindent Part (ii): Consider an environment with fundamental $\tilde{x}_\alpha$ and cost $\alpha \tilde{c}$. By definition,
\begin{equation*}
\overline{P}_{\alpha}(\emptyset)=\mathbb{E}(\alpha \tilde x| \alpha\tilde x- \alpha \tilde c<\overline{P}_{\alpha}(\emptyset)).
\end{equation*}
Multiplying both sides by $\frac{\alpha'}{\alpha}$, we get
\begin{equation*}
\frac{\alpha'}{\alpha} \overline{P}_{\alpha}(\emptyset)
=\mathbb{E}(\alpha' \tilde{x}| \alpha' \tilde{x}- \alpha' \tilde{c}<\frac{\alpha'}{\alpha} \overline{P}_{\alpha}(\emptyset)).
\end{equation*}
Thus $\frac{\alpha'}{\alpha} \overline{P}_{\alpha}(\emptyset)$ is an equilibrium non-disclosure price in the environment with fundamental $\alpha' \tilde x$ and cost $\alpha' \tilde{c}$. Moreover, the function $y \to \frac{\alpha'}{\alpha} y$ is a bijection of the non-disclosure price between the environment $(\alpha, \alpha \tilde{c})$ and the environment $(\alpha', \alpha' \tilde{c})$. So it maps
the maximal non-disclosure price $\overline{P}_{\alpha}(\emptyset)$ in $(\alpha, \alpha \tilde{c})$ to the maximal non-disclosure price $\frac{\alpha'}{\alpha} \overline{P}_{\alpha}(\emptyset)$ in
$(\alpha', \alpha' \tilde c)$.

Next, the disclosure event under $(\alpha', \alpha' \tilde{c})$ at price $\frac{\alpha'}{\alpha} \overline{P}_{\alpha}(\emptyset)$ is
\begin{equation*}
\alpha' \tilde{x}- \alpha' \tilde{c}\ \ge\ \frac{\alpha'}{\alpha} \overline{P}_{\alpha}(\emptyset)
\quad\Longleftrightarrow\quad
\alpha \tilde{x}- \alpha \tilde{c}\ \ge\ \overline{P}_{\alpha}(\emptyset).
\end{equation*}
Hence the disclosure probability is the same, i.e., $D_{(\alpha', \, \alpha' \tilde{c})}\ =\ D_{(\alpha, \, \alpha \tilde{c})}$. By assumption, holding the fundamental $\tilde x_{\alpha'}$ fixed, the cost $\alpha' \tilde{c}$ is higher than $\alpha \tilde{c}$. It then follows from part (i) that in the two environments with the same fundamental $\tilde x_{\alpha'} = \alpha' \tilde{x}$ but different
costs $\alpha' \tilde{c}$ and $\alpha \tilde{c}$, it must be that $D_{(\alpha', \, \alpha' \tilde{c})}\ \le\ D_{(\alpha', \, \alpha \tilde{c})}$. Combined with $D_{(\alpha', \, \alpha' \tilde{c})}\ =\ D_{(\alpha, \, \alpha \tilde{c})}$, it follows that $D_{\alpha} \equiv D_{(\alpha, \, \alpha \tilde{c})} \le D_{(\alpha', \, \alpha \tilde{c})} \equiv D_{\alpha'}$.
\qed

\textbf{Proof of Remark \ref{remark:fixed}:} 
Under fixed cost $c_0$, the manager discloses if and only if $x-c_0\ge P(\emptyset)$, i.e.,
$x\ge \tau$ where $\tau\equiv P(\emptyset)+c_0$. Further, $P(\emptyset)=\mathbb{E}(\tilde x| \tilde x< \tau)$ in the competitive market. Substituting $P(\emptyset)=\tau-c_0$ yields the equilibrium indifference condition
\begin{equation*}
\tau-c_0=\mathbb{E}(\tilde{x}| \tilde{x} < \tau)
\quad\Longleftrightarrow\quad
\delta(\tau)\equiv \tau-\mathbb{E}(\tilde{x}| \tilde{x}< \tau)=c_0,
\end{equation*}
where $\mathbb{E}(\tilde{x}| \tilde{x}< \tau) = \frac{\int_{-\infty}^\tau x d F(x)}{F(\tau)}$.

Moreover, item 1 follows because if $c_0 \ge \inf_{\tau: \,F(\tau)\in(0,1)}\delta(\tau)$ ($c_0 < \inf_{\tau: \,F(\tau)\in(0,1)}\delta(\tau)$), there exists (does not exist) $\tau \in \mathbb{R}$ such that~\eqref{eq:fixed_cost} holds. Item 2 holds because if $\delta(\tau)$ is strictly increasing in $\tau$,~\eqref{eq:fixed_cost} has at most one solution. Item 3 is immediate from item 1.
\qed

\textbf{Proof of Example \ref{ex:laplace}:}
The truncated means $\frac{\int_{-\infty}^\tau x d F(x)}{F(\tau)}$ of $\tilde{x}$ with density $f(x)=\frac{1}{2} e^{-|x|}$ are computed by standard integrations, which gives $\delta(\tau)$. Specifically, 
\begin{equation*}
    \mathbb{E}(\tilde{x}| \tilde{x} < \tau) = 
    \begin{cases}
    \tau - 1 & \text{ if } \tau \le 0 \\
\frac{\tau + 1}{1 - 2 e^\tau} & \text{ if } \tau > 0
    \end{cases}.
\end{equation*}
For $\tau>0$, differentiate $\delta(\tau)=\frac{2\tau+e^{-\tau}}{2-e^{-\tau}}$:
\begin{equation*}
\delta'(\tau) = \frac{(2-e^{-\tau})^2 - e^{-\tau}(2\tau+e^{-\tau})}{(2-e^{-\tau})^2} = \frac{4-2e^{-\tau}(2+\tau)}{(2-e^{-\tau})^2}
>0,
\end{equation*}
because $4-2e^{-\tau}(2+\tau) \bigg|_{\tau = 0} = 0$ and $4-2e^{-\tau}(2+\tau)$ is strictly increasing in $\tau > 0$. So $\delta(\tau)$ is strictly increasing in $\tau > 0$. Moreover, $\lim_{\tau\to 0}\delta(\tau)=1$ and $\lim_{\tau\to\infty}\delta(\tau)=\infty$.
Combined with $\delta(\tau)=1$ for all $\tau\le 0$, this implies: (i) if $c_0<1$, there is no solution to $\delta(\tau)=c_0$; (ii) if $c_0=1$, every $\tau\le 0$ is a solution; (iii) if $c_0>1$, there is a unique solution $\tau^*>0$.
\qed

\textbf{Proof of Lemma \ref{pd1}:} Equation (\ref{xx}) can be rewritten as follows:
\begin{eqnarray}
P^*(\emptyset)&=&\frac{\int \int_{-\infty}^{P^*(\emptyset)+c}x f(x)dx h(c)dc}{\int F(P^*(\emptyset)+c)h(c) dc}=  \frac{\int \int_{x-P^*(\emptyset)}^{\infty}x f(x) h(c)dcdx}{[F(P^*(\emptyset)+c)H(c)]^{\infty}_{-\infty}-\int f(P^*(\emptyset)+c)H(c) dc}\nonumber\\
&=& \frac{\int x f(x)(1-H(x-P^*(\emptyset)))dx}{1-\int f(P^*(\emptyset)+c)H(c) dc}= \frac{\mu-\int x f(x)H(x-P^*(\emptyset))dx}{1-\int f(x)H(x-P^*(\emptyset)) dx}\nonumber\\
\mu-P^*(\emptyset)&=& \int (x-P^*(\emptyset)) f(x)H(x-P^*(\emptyset)) dx = \int x f(x+P^*(\emptyset))H(x) dx, \label{fin0}
\end{eqnarray}
where the second equality is  a change in variable in the numerator and an integration by parts in the denominator, the third equality reorganizes terms, and the denominator of the fourth equality follows by changing variables after switching the labels from $c$ to $x$ as the integration variable. The sixth equality is another change of variables.
Rearranging (\ref{fin0}) further, $\Gamma(P^*(\emptyset))=0$, where 
\begin{eqnarray*}
\Gamma(P(\emptyset))&=&\mu-P(\emptyset)- \int x f(x+P(\emptyset))H(x) dx \nonumber\\
&=&\int x f(x+P(\emptyset)) dx- \int x f(x+P(\emptyset))H(x) dx \nonumber\\
&=&\int x f(x+P(\emptyset))(1-H(x)) dx.
\end{eqnarray*} \qed

\textbf{Proof of Proposition \ref{stat1}:}\ For the first part of the claim, note that if $H(0)=1$, equation (\ref{fin1}) equals
\begin{eqnarray*}
 \int x(1-H(x)) f(x+P(\emptyset)) dx=\int_{-\infty}^0 x (1-H(x)) f(x+P(\emptyset)) dx<0.
\end{eqnarray*}
For the second part of the claim, $P(\emptyset)<\mu$ follows immediately from (\ref{xx}).\ To show existence, consider the function $\Gamma(.)$ defined in Lemma \ref{pd1}.
Evaluating this function,
\begin{eqnarray*}
\Gamma(\mu)&=& \int x f(x+\mu)(1-H(x)) dx<(1-H(0))\int x f(x+\mu) dx=0.
\end{eqnarray*}
To complete the proof, it remains to be shown that $\Gamma(P(\emptyset))>0$ for $P(\emptyset)$ sufficiently small. By continuity, for some $c_0$ sufficiently small,
\begin{eqnarray*}
\Gamma(P(\emptyset))&=&\int x f(x+P(\emptyset))(1-H(x)) dx>\int^{c_0}_{-\infty} x f(x+P(\emptyset)) dx.
\end{eqnarray*}
where $\int^{c_0}_{-\infty} x f(x+P(\emptyset)) dx=0$ is the equilibrium with a fixed cost of $c_0$. Hence, it suffices to show that the model with fixed cost $c_0>0$ has an equilibrium. Recall that the equilibrium can be defined in terms of the following threshold equation: $\delta(\tau)\equiv \tau-\mathbb{E}(\tilde{x}|\tilde{x}< \tau)=c_0$.

We shall prove below that if $f(.)$ has a thin tail, 
\begin{equation}\label{eq:delta}
\liminf_{\tau\to -\infty}\delta(\tau)=0.
\end{equation}
Let
\begin{equation*}
\varphi(\tau)\equiv \int_{-\infty}^{\tau} f(t)(\tau-t)\,dt.
\end{equation*}
It follows from $\delta(\tau)\equiv \tau-\mathbb{E}(\tilde x| \tilde x< \tau)$ that $\delta(\tau)=\varphi(\tau)/F(\tau)$. Because $\varphi'(\tau)=\int_{-\infty}^{\tau} f(t)\,dt=F(\tau)$,
we have 
\begin{equation*}
\frac{1}{\delta(\tau)}=\frac{F(\tau)}{\varphi(\tau)}=\frac{\varphi'(\tau)}{\varphi(\tau)}=\bigl(\log \varphi(\tau)\bigr)'.
\end{equation*}
Consider any $a>0$. Let $M_a(\tau)\equiv \sup_{t\le \tau} e^{-at}f(t)$. Then for all $t\le \tau$, $f(t)\le M_a(\tau)e^{at}$. Further, if $\tilde{x}$ has a thin lower tail, i.e., $\lim_{x\to-\infty}e^{-ax}f(x)=0$, we have $M_a(\tau)\to 0$ as $\tau\to -\infty$. It follows that
\begin{equation*}
\varphi(\tau) \le M_a(\tau)\int_{-\infty}^{\tau} e^{at}(\tau-t)\,dt = M_a(\tau) \frac{e^{at}}{a^2} \left( a \left(\tau - t \right) + 1\right)\Big|_{-\infty}^{\tau} = 
M_a(\tau)\frac{e^{a\tau}}{a^2}.
\end{equation*}
Therefore, $e^{-a\tau}\varphi(\tau)\le M_a(\tau)/a^2\to 0$ as $\tau\to-\infty$.  
%%%
We now show that $\varphi'(\tau)/\varphi(\tau)$ cannot be bounded above on the lower tail.
Suppose by contradiction that there exist $\tau_0\in \mathbb{R}$ and $K>0$ such that
$\varphi'(t)/\varphi(t)\le K$ for all $t\le \tau_0$.
Then integrating the inequality from $\tau$ to $\tau_0$ gives
\begin{equation*}
\log \varphi(\tau_0)-\log \varphi(\tau) = \int_{\tau}^{\tau_0}\frac{\varphi'(t)}{\varphi(t)}\,dt \le K(\tau_0-\tau).
\end{equation*}
Hence, $\varphi(\tau)\ge \varphi(\tau_0)e^{-K\tau_0}e^{K\tau}$ for all $\tau\le\tau_0$.
Choose $a>K$. Then
\begin{equation*}
e^{-a\tau}\varphi(\tau)\ge \varphi(\tau_0)e^{-K\tau_0} e^{(K-a)\tau}\to \infty
\quad\text{as }\tau\to -\infty,
\end{equation*}
implies a contradiction to $e^{-a\tau}\varphi(\tau)\to 0$. Hence, for every $K>0$ and every $\tau_0$, there exists $\tau\le\tau_0$ such that
$\varphi'(\tau)/\varphi(\tau)>K$. So we have $1/\delta(\tau)>K$ and $\delta(\tau)<1/K$ as $\tau \to -\infty$, which proves \eqref{eq:delta}.
 
It follows that there is a solution $P_0$ to 
\begin{equation*}
 \int^{c_0}_{-\infty} x f(x+P_0) dx=0,
\end{equation*}
and this solution must also satisfy $\Gamma(P_0)>0$.

To show uniqueness, we differentiate \(\Gamma\) and  obtain 
\[
\Gamma'(P) 
\;
\;=\;
\int_{-\infty}^{\infty}  x(1-H(x)) f'(x+P)\,dx.
\]
Using the logarithmic derivative \(\ell_P(x) = \frac{f'(x+P)}{f(x+P)} = (\log f)'(x+P)\), we can write
\[
\Gamma'(P)
\;=\;
\int_{-\infty}^{\infty}  x(1-H(x))\,\ell_P(x)\,f(x+P)\,dx.
\]
By the monotone likelihood ratio property, \(\ell_P(x)\) is a non-increasing function of \(x\). Define \[
w_P(x) \equiv (1-H(x))\,f(x+P).
\]
We normalize \(w_P(x)\) to obtain a probability density:
\[
\pi_P(x) = \frac{w_P(x)}{W_P},
\quad\text{where}\quad
W_P = \int_{-\infty}^{\infty} w_P(x)\,dx > 0.
\]
Given that at a certain \(P\) we have:
\[
\Gamma(P) = \int_{-\infty}^{\infty} x\,w_P(x)\,dx = 0.
\]
It follows that\[
\mathbb{E}_{\pi_P}(x) = \int_{-\infty}^{\infty} x\,\pi_P(x)\,dx = 0
\quad \text{ and } \quad
\Gamma'(P)
\;=W_{P}\;
\int_{-\infty}^{\infty}  x\,\ell_P(x)\,\,\pi_P(x)\,\,dx.
\]
By the Chebyshev integral inequality for monotone functions of opposite sense, the covariance
\[
\operatorname{Cov}_{\pi_p}\big(x,\ell_P(x)\big) \;<\; 0.
\]
Further, $\operatorname{Cov}_{\pi_p}\big(x,\ell_P(x)\big)=\mathbb{E}_{\pi_P}(x \ell_P(x))$ because $\mathbb{E}_{\pi_P}(x) =0$. Hence, $
\Gamma'(P) < 0$ for all $P\in\mathbb{R}.$ That is, \(\Gamma(P)\) is strictly decreasing in \(P\).
\qed

\textbf{Proof of\ Proposition \ref{rofo}:} 
Let $\Phi(.)$ be the c.d.f.\ and $\phi(.)$ the p.d.f.\ of a standard normal random variable. These comparative statics follow from the implicit function theorem, given that, due to the fact that $\Gamma'(P(\emptyset))<0$, the sign of $\partial P(\emptyset)/\partial \gamma$ is the same as the sign of $\partial \Gamma/\partial \gamma$
\begin{equation*}
\frac{\partial \Gamma}{\partial m}=\frac{1}{\sigma}\int x \phi(\frac{x-m}{\sigma}) f(x+P(\emptyset))dx.
\end{equation*}
Define
\begin{equation*}
w(x) \equiv [1-\Phi({\textstyle\frac{x-m}{\sigma}})]f(x+P(\emptyset)) \quad \text{ and } \quad
W \equiv \int_{-\infty}^{\infty} w(x)dx>0,
\end{equation*}
such that $g(x) \equiv \frac{w(x)}{W}$ is a p.d.f. with mean $\mathbb{E}_g(x)=\int x g(x)dx=0$ at an equilibrium $P(\emptyset)$. Defining the (increasing) inverse Mills ratio as $\rho(x) \equiv \frac{\phi(x)}{1-\Phi(x)}$, we get
\begin{equation*}
\frac{\partial \Gamma}{\partial m}=\frac{W}{\sigma}\int_{-\infty}^{\infty}x\rho\bigl({\frac{x-m}{\sigma}}\bigr)\,g(x)\,dx
    =\frac{W}{\sigma}\mathbb{E}_g\bigg(x\,\rho(\frac{x-m}{\sigma})\bigg)
  = \frac{W}{\sigma} \operatorname{Cov}_g(x,\rho(\frac{x-m}{\sigma}))>0.
\end{equation*}
Similarly,
\begin{equation*}
\frac{\partial \Gamma}{\partial \sigma}=\frac{1}{\sigma^2}\int x(x-m) \phi(\frac{x-m}{\sigma})f(x+P(\emptyset))dx
  = \frac{W}{\sigma} \operatorname{Cov}_g(x,(x-m)\rho(\frac{x-m}{\sigma})),
\end{equation*}
where $(x-m)\rho(\frac{x-m}{\sigma})$ is increasing in $x$ over a set of realizations whose probability converges to one, for $m$ sufficiently small or $\sigma$ sufficiently large, implying that this term is positive.
\qed

\end{singlespace}

%\bibliographystyle{cje}
%\bibliography{general5}
\ifx\undefined\bysame
\newcommand{\bysame}{\leavevmode\hbox to\leftmargin{\hrulefill\,\,}}
\fi

%\bibliography{../../biblio/general5}

\end{document}